\documentclass[
10pt,
aps,
prd,
preprintnumbers,
superscriptaddress,
showpacs,
showkeys,
floatfix,
amsmath,
amssymb,
amsfonts,
twocolumn,
nofootinbib,
]{revtex4-2}

\usepackage{graphicx}
\usepackage{xcolor}
\usepackage{booktabs}
\usepackage{amsthm,amscd}
\usepackage{enumitem}
\usepackage{dcolumn}

\usepackage{silence}
\usepackage{hyperref}

\makeatletter
\def\footnoterule{\kern-3\p@
  \hrule \@width 2in \kern 2.6\p@
  \kern 1\p@}
\makeatother

\begin{document}


    \title{Diagonal Kenney-Laub Rational Approximation to the Overlap Operator using Wilson and Brillouin Kernel}
    
    \author{Stephan D\"urr}
    \affiliation{University of Wuppertal, Gaußstraße 20, 42119 Wuppertal, Germany}
    \affiliation{JSC, Forschungszentrum J\"ulich, 52425, J\"ulich, Germany}
      
    \author{Stylianos Gregoriou}
    \author{Giannis Koutsou}
    \affiliation{The Cyprus Institute, CaSToRC, 20 Kavafi Street, Nicosia 2121, Cyprus}
    
    \begin{abstract}
        We propose a formulation of the overlap Dirac operator in lattice QCD that employs diagonal Kenney-Laub (KL) iterates to approximate the matrix sign function. KL iterates require no prior spectral information about the kernel operator and, when expressed via their partial fraction decomposition, offer a practical and efficient approximation scheme. We evaluate this approach in a proof-of-concept implementation using quenched lattices at $\beta=6.2$ and two Dirac operator discretizations as a kernel, namely the Wilson and the Brillouin operators. By examining the approximate overlap operator's violation of the Ginsparg-Wilson relation and the critical bare quark mass for increasing approximation order, we find that KL iterates deliver enhanced chiral symmetry preservation and computational efficiency compared to the Chebyshev polynomial approach.
    \end{abstract}
    
    \maketitle 
    


    \section{Introduction}
    \label{sec:intro}
        
        Practical implementations of the overlap operator, a lattice discretization of the QCD Dirac operator~\cite{Neuberger:1997fp, Neuberger:1998wv} that preserves the chiral properties of the continuum theory, have posed significant computational challenges since its inception~\cite{Neuberger:1998my}, both in terms of construction and cost-effectiveness. Various approaches have been developed over the years to enhance the efficiency and applicability of the overlap procedure. In this work, we propose a specific method that balances algorithmic simplicity with computational effectiveness.
        
        \medskip
        
        Despite their considerable computational cost, overlap fermions offer significant advantages in lattice QCD studies where chiral symmetry is essential. Their close adherence to the Ginsparg–Wilson relation~\cite{Ginsparg:1981bj, Luscher:1998pqa} makes them well-suited for precision studies of topological phenomena in QCD~\cite{DelDebbio:2003rn}, improved determinations of quark masses and decay constants~\cite{Boyle:2017jwu}, as well as investigations of the QCD vacuum structure~\cite{Luscher:1996sc}, where traditional formulations suffer from increased chiral symmetry breaking.
        
        Numerous techniques have been proposed to improve the efficiency of the overlap operator, such as polynomial and rational approximations of the matrix sign function~\cite{Edwards:1998yw, vandenEshof:2002ms, Kennedy:2004tj}, improved kernel choices~\cite{DeGrand:2000tf, Bietenholz:2002ks, Ikeda:2009mv, Durr:2010ch, Cho:2015ffa}, and optimization of solver strategies~\cite{Alexandru:2011sc, Brannick:2014vda}. While many of these approaches succeed in improving performance, they often introduce substantial complexity in implementation or require careful tuning of multiple parameters.
        
        \medskip
        
        In this paper, we investigate an approach for constructing the overlap operator that combines three ingredients. First, we employ the \emph{diagonal Kenney-Laub (KL) iterates}, a globally convergent rational iterative scheme for approximating the matrix sign function that can be systematically improved by increasing the approximation order. In our implementation, no estimate of the extreme eigenvalues of the kernel operator is required, thereby reducing computational complexity.
        
        Second, we express the KL iterates via their \emph{partial fraction decomposition}, with constants that depend only on the approximation order and are calculated from simple trigonometric expressions. This decomposition enables the use of robust solvers such as the Multi-Shift Conjugate Gradient (MSCG) algorithm, which requires minimal parameter tuning.
        
        Third, we adopt the \emph{Brillouin operator} as the kernel of the overlap procedure, a parameter-free improved Wilson-like operator that naturally exhibits good chiral properties in terms of how well it satisfies the Ginsparg-Wilson relation. The goal of this paper is to provide a detailed assessment of this combination and demonstrate its practical viability as an implementation of the overlap operator in a vector-like gauge theory, such as QCD.
        
        \medskip
        
        The Brillouin-improved Wilson-Dirac operator, or Brillouin operator, introduced in Ref.~\cite{Durr:2010ch} (see Appendix \ref{app:Brillouin} for definition), was found to have a spectrum that resembled that of the ideal Ginsparg-Wilson (GW) fermions more closely than the Wilson-Dirac operator. In later work~\cite{Durr:2017wfi}, we explored the use of the diagonal KL iterates (reviewed in Appendix \ref{app:KL}) for approximating the matrix sign function and observed promising numerical behavior. A preliminary application of these iterates in overlap operator applications also yielded encouraging results in terms of inversion cost and residual mass~\cite{Durr:2025cxn}.
        
        Here, we extend these investigations by combining the diagonal KL rational approximation with the Brillouin kernel. Moreover, the incorporation of the former's partial fraction decomposition, as outlined above, enables us to reach higher orders of the approximation than was previously carried out in Ref.~\cite{Durr:2017wfi}. We benchmark this approach against the well-established Chebyshev polynomial approximation~\cite{vandenEshof:2002ms}, reviewed in Appendix~\ref{app:Chebyshev}.
        
        \medskip
        
        The remainder of this paper is structured as follows. In Sec.~\ref{sec:overlap_intro}, we introduce the overlap operator formulation and the partial fraction decomposition of KL iterates. In Sec.~\ref{sec:overlap_implementation}, we describe our numerical setup and quantify the Ginsparg-Wilson violation as a function of the diagonal KL order, for both the Wilson and Brillouin kernels. In Sec.~\ref{sec:KL_observables}, we compute the critical bare quark mass as a function of the diagonal KL order, again for both kernels. In Sec.~\ref{sec:comparison}, results obtained using KL iterates are compared to those using the Chebyshev polynomial approximation, including a computational cost comparison across all method-kernel combinations. In Sec.~\ref{sec:conclusions} we summarize our conclusions. Auxiliary information is provided in four appendices: Appendix~\ref{app:implementation} gives details of the iterative solver implementations and precision choices; Appendix~\ref{app:Brillouin} summarizes the construction and properties of the Brillouin operator; Appendix~\ref{app:KL} presents the Kenney-Laub rational approximation, with particular focus on the diagonal iterates and their partial fraction decomposition; and Appendix~\ref{app:Chebyshev} outlines the Chebyshev polynomial approximation as implemented here.
    

    \section{Overlap Operator}
    \label{sec:overlap_intro}
        
        \subsection{Formulation}
            
            Following the conventions introduced in Ref.~\cite{Durr:2017wfi}, the (dimensionless) massive overlap operator at the bare quark mass $m$ is defined as:
            \begin{align}
                a D^{\mathrm{ov}}_m = \left( \rho + \frac{a m}{2} \right) \mathbb{I} + \left(\rho - \frac{a m}{2}\right)\! \gamma_5 \, \textrm{sgn}\!\left[ \mathbb{X} \right] \; , 
                \label{eq:massive_overlap_definition}
            \end{align}
            where $a$ is the lattice spacing and the argument of the matrix sign function is given by:
            \begin{align}
                \mathbb{X} = \gamma_5 \, a D^{\text{ker}}_{-\rho/a}
                \equiv \gamma_5 \! \left( a D^{\text{ker}}_{0} - \rho \mathbb{I} \right) \; .
                \label{eq:gamma5_kernel}
            \end{align}
            Here, $\mathbb{I}$ is the identity matrix, $D^{\text{ker}}_{0}$ is a massless doubler-free Dirac operator serving as the \emph{kernel} of the overlap operator, and $\rho$ with $0 < \rho < 2$ is a dimensionless parameter that serves as the kernel's bare quark mass. Setting the bare overlap operator's mass to zero ($m=0$), Eq.~\ref{eq:massive_overlap_definition} simplifies to the massless overlap operator%
            \footnote{The massive overlap operator can alternatively be understood as the result of adding a chirally-rotated mass term to the massless operator: $a D^{\mathrm{ov}}_m = a D^{\mathrm{ov}} + a m\!\left( \mathbb{I} - \frac{1}{2 \rho} a D^{\mathrm{ov}} \right)$.}:
            \begin{align}
                a D^{\mathrm{ov}} \equiv \rho  \left( \mathbb{I} + \gamma_5 \, \textrm{sgn}\!\left[ \mathbb{X} \right] \right) \; .
                \label{eq:massless_overlap_definition}
            \end{align}
            Since $\mathbb{X}$ is Hermitian by construction, we restrict our focus to Hermitian arguments of the matrix sign function throughout this work.
            
            The matrix sign function can be defined for any Hermitian matrix $\mathbb{H}$, with positive-definite $\mathbb{H}^2$, as:
            \begin{align}
                \text{sgn}[\mathbb{H}] = \mathbb{H} \left( \mathbb{H}^{\dagger} \mathbb{H} \right)^{-1/2}
                = \mathbb{H} \left( \mathbb{H}^2 \right)^{-1/2} \; .
                \label{eq:matrix_sign_function_definition}
            \end{align}
            This matrix is also Hermitian ($\text{sgn}^{\dagger}[\mathbb{H}] = \text{sgn}[\mathbb{H}]$) and unitary:
            \begin{align}
                \text{sgn}^{\dagger}[\mathbb{H}] \text{sgn}[\mathbb{H}] = \text{sgn}^2[\mathbb{H}] = \mathbb{I} \; .
                \label{eq:sign_squared_property}
            \end{align}
            
        \subsection{Sign function approximation}
    
            A main component of this work is the use of Kenney-Laub (KL) iterates to approximate the sign function. In our previous work~\cite{Durr:2017wfi}, we introduced the KL family of iterations, emphasizing the diagonal $(n,n)$ iterates and demonstrating their advantageous properties in terms of convergence behavior and numerical stability (see Appendix~\ref{app:KL} for details). Furthermore, when combined with the Brillouin kernel, we found that KL iterates also provide improved localization and normality properties, and naturally facilitate cascaded preconditioning strategies through a sequential application of different rational approximations.
    
            All calculations in that prior investigation employed the original form of the diagonal KL iterates, expressing the matrix sign function approximation as a single Pad\'{e} approximant fraction%
            \footnote{See Tables 1 and 2 of~\cite{Durr:2017wfi} for examples of this formulation for several $n$.  } (see Eq.~\ref{eq:KL_iterates_single_fraction}):
            \begin{align}
                \textrm{sgn}\!\left[ \mathbb{X} \right] \approx \mathbb{X} \frac{ \sum_{i=0}^{n} \binom{2n+1}{2i+1} ( \mathbb{X}^{2} )^i }{ \sum_{i=0}^{n} \binom{2n+1}{2i} ( \mathbb{X}^{2} )^i }
                \equiv \mathbb{X} \frac{P_{nn}(\mathbb{X}^2)}{Q_{nn}(\mathbb{X}^2)} \; .
                \label{eq:KL_sign_function_single_fraction}
            \end{align}
            We refer to this form as the ``single-fraction'' (SF) expression. In this form, both the numerator and denominator polynomials are of degree $2n$, and the inverse of $Q_{nn}(\mathbb{X}^2)$ is required. However, since the condition number of $Q_{nn}(\mathbb{X}^2)$ grows rapidly with $n$, this inversion becomes prohibitively costly at higher orders. As a result, in the investigation of KL iterates carried out in Ref.~\cite{Durr:2017wfi}, we were limited to diagonal KL orders $n = 1$ and $n=2$.
    
            However, as with any rational function, the diagonal KL iterates can alternatively be expressed via a partial fraction (PF) decomposition. For the diagonal KL iterates, the decomposition takes the numerically convenient form (see Eq.~\ref{eq:KL_sign_function_partial_fraction}):
            \begin{align}
                \textrm{sgn}\!\left[ \mathbb{X} \right] \approx \mathbb{X} \! \left( c_0 + \sum_{i=1}^{n} \frac{c_i}{ \mathbb{X}^2 + \sigma_i } \right) \; ,
            \end{align}
            with coefficients and shifts determined solely by the diagonal KL order $n$ through simple trigonometric expressions (Eqs.~\ref{eq:KL_partia_fraction_shifts} and \ref{eq:KL_partia_fraction_const}). In this formulation, the diagonal KL order $n$ coincides with the number of fractions in the expression.
    
            This approach is particularly well-suited for implementation with a Multi-Shift Conjugate Gradient (MSCG) solver~\cite{Frommer:1995ik, Jegerlehner:1996pm}, which enables the simultaneous inversion of all terms of the form $\mathbb{X}^2 + \sigma_i$. This significantly enhances computational efficiency, as the overall cost is largely governed by the fraction with the smallest shift, which converges the slowest.
    
            It should be noted that Neuberger originally suggested such an implementation in Ref.~\cite{Neuberger:1998my}. In particular, the following family of iterations for the sign function was proposed:
            \begin{align}
                f_n(z)=\frac{(1+z)^{2 n}-(1-z)^{2 n}}{(1+z)^{2 n}+(1-z)^{2 n}},
            \end{align}
            which, in KL terminology, correspond to the $(n-1,n)$ iterates. Both these iterates and the $(n,n)$ iterates proposed here are classified as Principal Pad\`{e} iterations~\cite{kenney1994hyperbolic-364}. To our knowledge, there has been no complete and systematic study that includes practical implementations of these iterates to date.
    
    
    \section{Overlap Operator Implementation}
    \label{sec:overlap_implementation}
    
        \subsection{Setup}
        \label{sec:setup}
    
            We evaluate our approach on an ensemble of quenched gauge configurations with lattice size $48 \times 24^3$ and $\beta = 6.2$, which corresponds to a lattice spacing of $a\simeq 0.067$~fm. The configurations are smeared with one iteration of APE smearing~\cite{APE:1987ehd} with $\alpha_{\text{APE}} = 0.72$.
    
            When using either the Wilson or Brillouin operators as a kernel to the overlap, we set $c_\mathrm{SW} = 0$ in Eq.~\ref{eq:Dirac_operator_structure}, and the dimensionless parameter $\rho$ in Eq.~\ref{eq:massive_overlap_definition} is set to unity ($\rho = 1$). This choice maintains consistency with our previous investigations~\cite{Durr:2010ch,  Durr:2017wfi}.
    
        \subsection{Comparison of partial and single fraction approaches}
                
            In the numerical tests that follow, we will use the partial fraction (PF) decomposition of the diagonal KL iterates. For completeness, we provide a comparison of the resources required to apply the overlap operator for the same diagonal KL order using either the PF or the single fraction (SF) approach. Namely, in Table~\ref{tab:partial_Vs_single_fraction} we provide the average core-hour cost for a single forward application of the massless overlap operator on a Gaussian random vector. For details regarding the implementation of the Multi-Shift Conjugate Gradient (MSCG) algorithm for the PF case, we refer to Appendix~\ref{app:implementation}. Note that for a fair comparison, the stopping criterion is kept fixed as $n$ increases. The comparison in Table~\ref{tab:partial_Vs_single_fraction} clearly demonstrates the significant advantage of using the PF expression for approximating the matrix sign function. Unless otherwise stated, the use of the partial fraction decomposition will be implied when referring to KL iterates from here on.
            
            \begin{table}[ht]
                \centering
                \caption{Average computational cost in core-hours per random vector for the forward application of the massless overlap operator.  We compare the single-fraction (SF) and partial-fraction (PF) formulations of the diagonal KL orders $n = 1, 2, \dots, 5$ for the Wilson and Brillouin kernels. The comparison is carried out on a quenched gauge field configuration with the parameters given in Sec.~\ref{sec:setup}.}
                \label{tab:partial_Vs_single_fraction}
                
                \setlength{\tabcolsep}{8pt}
                \begin{tabular}{c| c c c c c}
                    \toprule
                    \multicolumn{6}{c}{Wilson Kernel} \\
                    \cmidrule(r){1-6}
                    $n$ & $1$ & $2$ & $3$ & $4$ & $5$ \\
                    \midrule
                    SF & 0.091 & 1.22 & 12.6 & 118.0 & $>$384 \\
                    PF & 0.072 & 0.122 & 0.162 & 0.195 & 0.221 \\
                    \toprule
                    \multicolumn{6}{c}{Brillouin Kernel} \\
                    \cmidrule(r){1-6}
                    $n$ & $1$ & $2$ & $3$ & $4$ & $5$ \\
                    \midrule
                    SF & 0.078 & 0.293 & 0.834 & 2.13 & 4.80 \\
                    PF & 0.071 & 0.116 & 0.155 & 0.183 & 0.203 \\
                    \bottomrule
                \end{tabular}
            \end{table}
    
        \subsection{Ginsparg-Wilson relation and sign function convergence}
        \label{sec:KL_forward}
            
            To assess the chiral properties of the overlap operator constructed using diagonal KL iterates, we investigate the degree to which the massless overlap operator in Eq.~\ref{eq:massless_overlap_definition} satisfies the Ginsparg-Wilson  relation:
            \begin{gather}
              \gamma_5 a D^{\mathrm{ov}} + a D^{\mathrm{ov}} \gamma_5 = \frac{1}{\rho} a D^{\mathrm{ov}} \gamma_5 a D^{\mathrm{ov}}
              \label{eq:Ginsparg_Wilson_relation}\;.
            \end{gather}
            With the diagonal KL order $n$ as the primary control parameter, we essentially examine how increasing $n$ changes the degree to which the GW relation is satisfied.
            
            To this end, as also observed in Ref.~\cite{Cundy:2010uq}, for the massless overlap operator, it is sufficient to test how well the matrix sign
            function is approximated. Substituting Eq.~\ref{eq:massless_overlap_definition} into Eq.~\ref{eq:Ginsparg_Wilson_relation} and taking the difference
            $\Delta_{\textrm{GW}}$ between the left and right sides yields:
            \begin{align}
                \Delta_{\textrm{GW}} = \rho \gamma_5 \left( \textrm{sgn}^2\!\left[ \mathbb{X} \right] - \mathbb{I} \right) \equiv \rho \gamma_5 \Delta_{\mathrm{sgn}^2}
            \end{align} 
            where $\Delta_{\mathrm{sgn}^2} = \textrm{sgn}^2\!\left[ \mathbb{X} \right] - \mathbb{I}$. This shows that an approximation satisfying the ``sign-squared'' property (Eq.~\ref{eq:sign_squared_property}) automatically satisfies the GW relation.
                    
            We therefore quantify the matrix sign function convergence using the ``sign-squared violation'' metric:
            \begin{align}
              \delta_{\mathrm{sgn}^2} = \frac{|| \Delta_{\mathrm{sgn}^2} \eta||^2}{||\eta||^2} = \frac{|| \left(\textrm{sgn}^2\!\left[ \mathbb{X} \right] - \mathbb{I} \right)\!\eta||^2}{||\eta||^2}
              \label{eq:sign_squared_violation}
            \end{align}
            which represents the relative squared norm of $\Delta_{\mathrm{sgn}^2}$ acting on a random vector $\eta$. The corresponding ``GW violation'' metric, $\delta_{GW} \equiv ||\Delta_{\textrm{GW}}\eta||^2/||\eta||^2$, is related by $\delta_{GW} = \rho^2 \delta_{\mathrm{sgn}^2}$. For our choice of $\rho = 1$, these metrics are identical: $\delta_{GW} = \delta_{\mathrm{sgn}^2}$.
            
            \medskip
            
            To systematically study the sign-squared violation, we select 12 gauge field configurations for which the operator $\mathbb{X}^2$ has condition numbers ($\kappa_{\mathbb{X}^2}$) that span the entire range of condition numbers of the ensemble, from approximately $10^2$ to $10^7$. Note that the condition number of $\mathbb{X}^2$ is a direct indicator of convergence behavior, since this is the operator that enters the MSCG.
            
            \begin{figure*}[ht]
                \centering
                \begin{minipage}{0.49\textwidth}
                    \centering
                    \includegraphics[width=\linewidth]{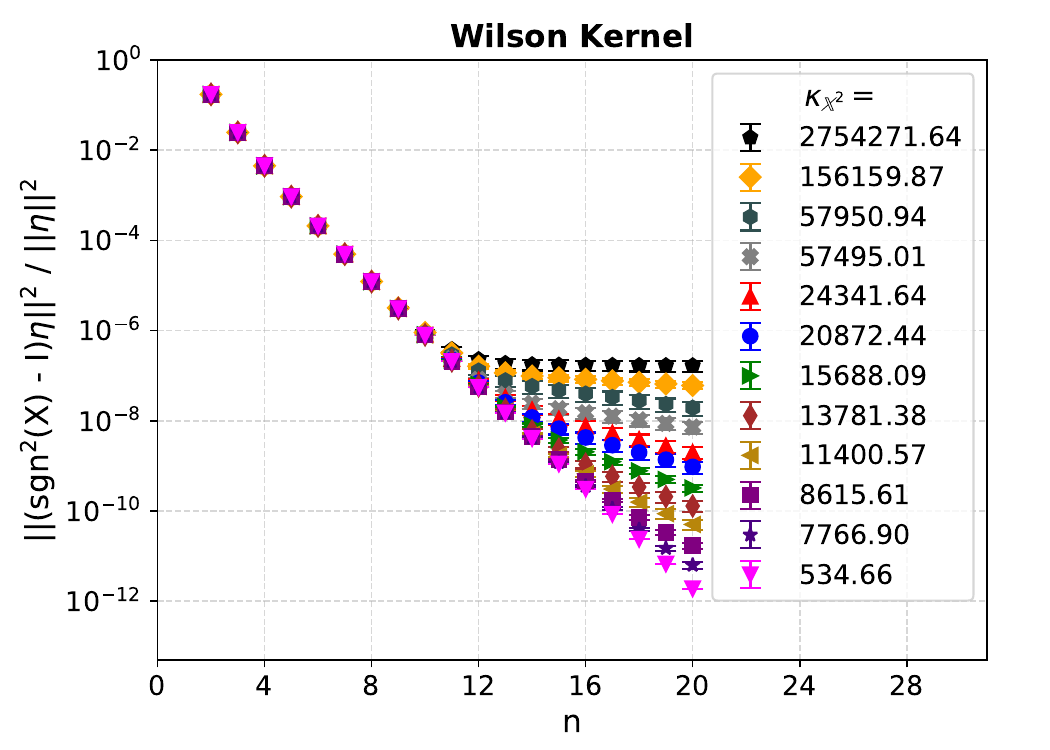}
                \end{minipage}
                \begin{minipage}{0.49\textwidth}
                    \centering
                    \includegraphics[width=\linewidth]{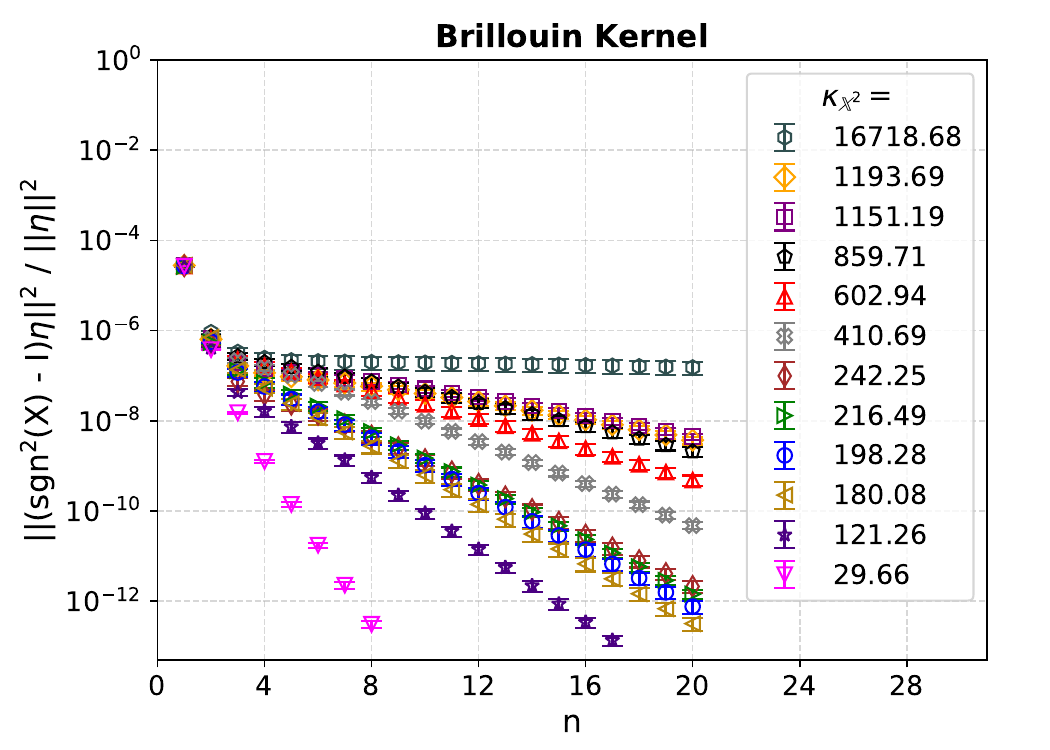}
                \end{minipage}
        
                \caption{Sign-squared violation $\delta_{\mathrm{sgn}^2}$ as a function of the diagonal Kenney-Laub order ($n=1, 2, \ldots, 20$) at constant $\epsilon_{\text{MSCG}} = 10^{-7}$ for 12 selected gauge field configurations, using the Wilson (left panel, filled markers) and the Brillouin (right panel, empty markers) kernel. The average and error for each case is obtained using 10 random vectors ($\eta$). The same color and symbol are used to identify any given configuration in either panel, while the condition number ($\kappa_{\mathbb{X}^2}$) is given for each kernel in the legend.}
                \label{fig:KL_sign_squared_Vs_n_by_config}
            \end{figure*}
            
            \begin{figure*}[ht]
                \centering
                \begin{minipage}{0.49\textwidth}
                    \centering
                    \includegraphics[width=\linewidth]{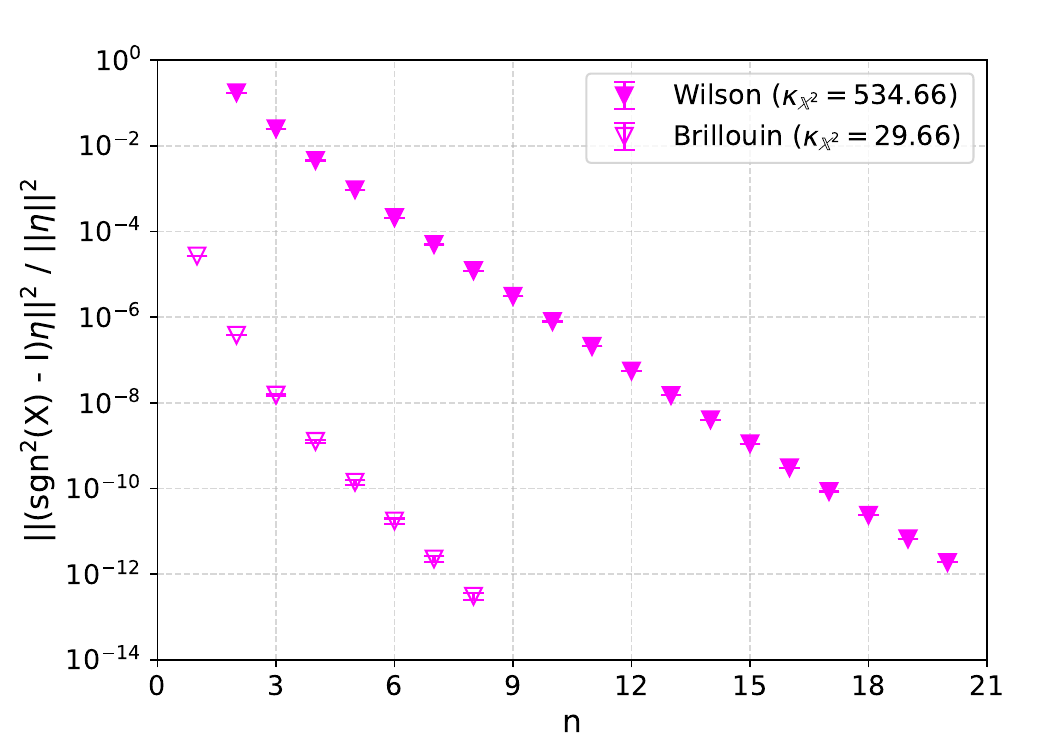}
                \end{minipage}
                \begin{minipage}{0.49\textwidth}
                    \centering
                    \includegraphics[width=\linewidth]{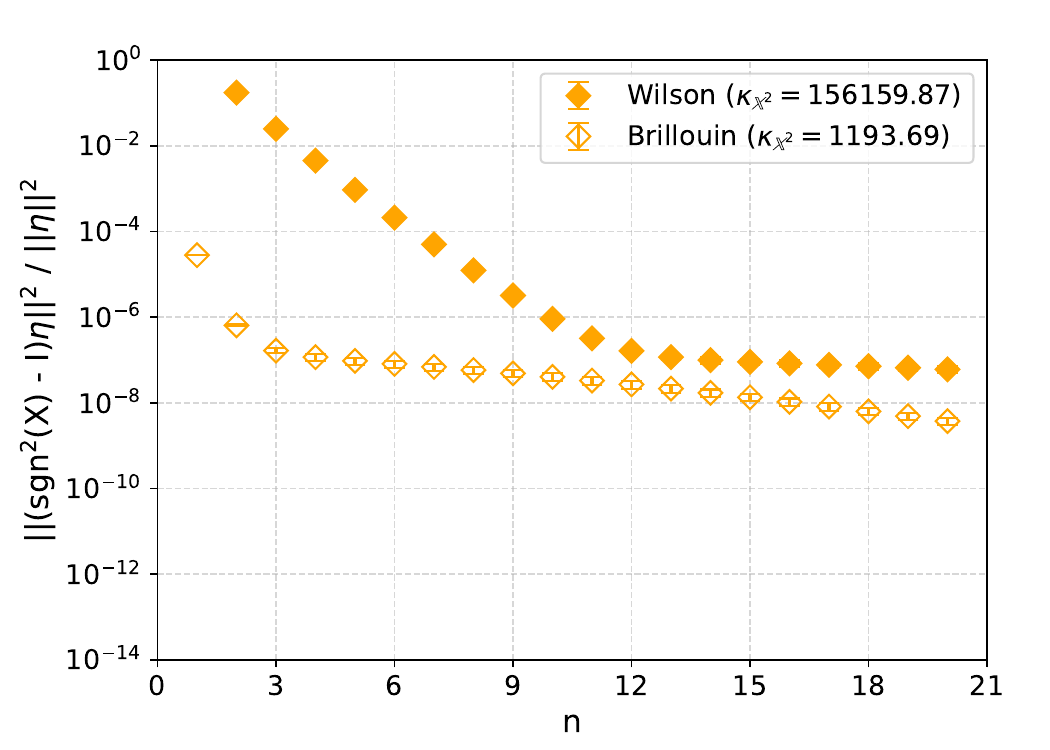}
                \end{minipage}
                
                \caption{Sign-squared violation $\delta_{\mathrm{sgn}^2}$ as a function of the diagonal Kenney-Laub order ($n=1, 2, \dots, 20$) for a relatively well-conditioned (left) and worse-conditioned (right) gauge configuration, for the Wilson and Brillouin kernels (full/open symbols).}
                \label{fig:KL_sign_squared_Vs_n_by_kernel}
            \end{figure*}
            
            In Fig.~\ref{fig:KL_sign_squared_Vs_n_by_config}, we plot the sign-squared violation for both Wilson and Brillouin kernels as a function of the diagonal KL order $n$, for $n=1,2,\dots,20$, where each data point represents an average over 10 random vectors. The results demonstrate that increasing the diagonal KL order provides systematic, monotonic improvement in the sign function approximation, even for poorly conditioned configurations.
    
            We note that the ordering of the configurations in terms of the magnitude of their condition numbers differs between the Wilson and Brillouin kernels. This is unsurprising, since no simple relation exists between $\kappa_{\mathbb{X}^2}$ of the two kernels. Furthermore, we observe that for a given configuration, $\kappa_{\mathbb{X}^2}$ tends to be larger for the Wilson kernel compared to Brillouin.
    
            Another observation is that for small $n$, all configurations exhibit a similar dependence on $n$ regardless of their condition number. However, after approximately $\delta_{\mathrm{sgn}^2}\simeq 5 \times 10^{-7}$, the sign-squared violation dependence on $n$ begins to spread more significantly, with the convergence rate becoming increasingly dependent on $\kappa_{\mathbb{X}^2}$.
    
            A possible explanation for this behavior is that the sign function approximation error is concentrated differently for different KL orders~\cite{Cundy:2010uq} (see Fig.~\ref{fig:scalar_sign_function_approximations}). Namely, at lower $n$, the approximation error is concentrated, in absolute value, near the upper end of the eigenvalue spectrum of the kernel operator (see Appendix~\ref{app:Brillouin} for the relevant bounds for both kernels), which varies little between configurations, leading to a universal convergence behavior; at higher orders, the error becomes concentrated near zero, where the eigenvalue distribution differs drastically across configurations. This may cause the observed spread of the convergence rates.
            
            \begin{figure*}[ht]
                \centering
                \begin{minipage}{0.49\textwidth}
                    \centering
                    \includegraphics[width=\linewidth]{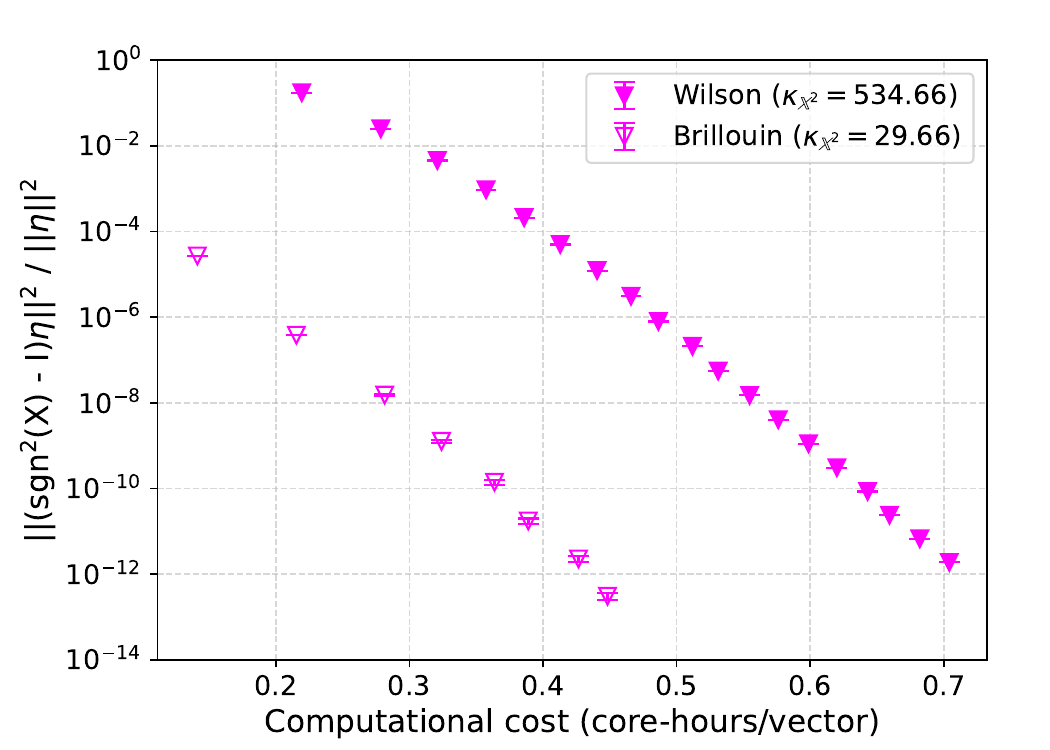}
                \end{minipage}
                \begin{minipage}{0.49\textwidth}
                    \centering
                    \includegraphics[width=\linewidth]{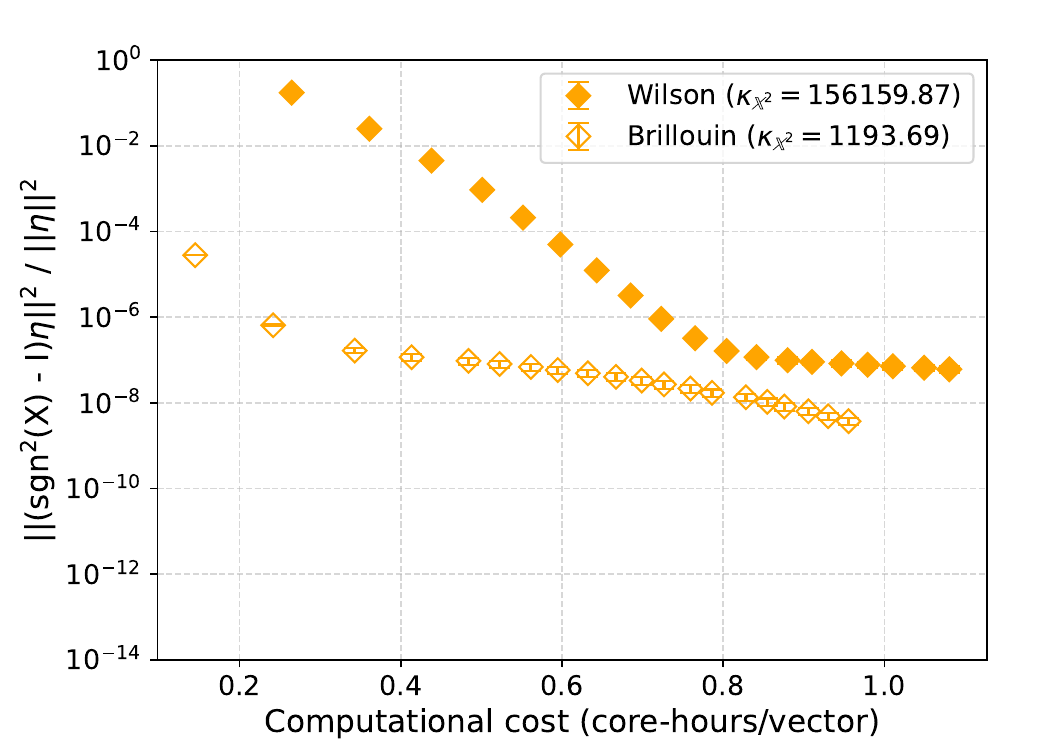}
                \end{minipage}
            
                \caption{Same data as in Fig.~\ref{fig:KL_sign_squared_Vs_n_by_kernel}, plotted against the average computational cost (in core-hours per random vector).}
                \label{fig:KL_sign_squared_Vs_core_hours}
            \end{figure*}
                    
            For a more direct comparison between the two kernels, in Fig.~\ref{fig:KL_sign_squared_Vs_n_by_kernel}, we plot the sign-squared violation versus $n$ for one ``well-conditioned'' configuration, with relatively small $\kappa_{\mathbb{X}^2}$, and one ``worse-conditioned'' configuration. The smaller condition numbers, typically associated with the Brillouin kernel, lead to faster convergence, requiring a lower diagonal KL order $n$ to achieve a given sign-squared violation $\delta_{\mathrm{sgn}^2}$ threshold.  However, for poorly conditioned configurations, the Brillouin kernel may exhibit shallower convergence with the spreading starting at smaller $n$ values. This can occasionally result in the Wilson kernel performing comparably to, or even gaining an advantage at larger $n$ values, as suggested in the right panel of Fig.~\ref{fig:KL_sign_squared_Vs_n_by_kernel}. Identifying optimal $n$ ranges for various precision targets is of interest to this study.
            
            Given that the Brillouin kernel has a higher computational cost (per forward application) than the Wilson kernel (see Appendix \ref{app:Brillouin}), it is worth investigating the computational efficiency of either operator as a kernel in the sign function. In Fig.~\ref{fig:KL_sign_squared_Vs_core_hours}, we show the sign-squared violation against average computational cost in core-hours per random vector for the same two configurations from Fig.~\ref{fig:KL_sign_squared_Vs_n_by_kernel}. These results demonstrate that (on the vast majority of configurations) the Brillouin kernel enables a more economical calculation of the overlap operator for a given $\delta_{\mathrm{sgn}^2}$ threshold. In other words, it seems that the reduced number of forward applications of the shifted Brillouin kernel (due to its better condition number) outweighs its higher per-application cost compared to the shifted Wilson kernel.
    
    
    \section{Physical observables and critical bare mass}
    \label{sec:KL_observables}
    
        We now examine the efficiency of the proposed overlap formulation in terms of quantities that require inverting the overlap operator, rather than the forward applications investigated so far. In particular, we compute the Partially Conserved Axial Current (PCAC) quark mass $am_\mathrm{PCAC}(t)$ and the pion effective mass $aM^\mathrm{eff}_{\pi}(t)$, and investigate their dependence on the bare overlap quark mass and the diagonal KL order, extending the analysis from our previous work~\cite{Durr:2017wfi}.
            
            \begin{table}[!ht]
                \centering
                \caption{Average computational cost (in core-hours per spinor) for inverting the massive overlap operator (via CGNR) for $am=0.15$ at diagonal KL orders $n = 1, 2, 3, 4$, using three different formulations: partial-fraction (PF), multiply-up (MU) trick of Eq.~\ref{eq:multiply_up_trick}, and single-fraction (SF). Results are shown for the Wilson and the Brillouin kernels, using the same gauge field configuration as in Table~\ref{tab:partial_Vs_single_fraction}.}
                \label{tab:partial_Vs_single_fraction_invert}
    
                \setlength{\tabcolsep}{10pt}
                \begin{tabular}{c| c c c c}
                    \toprule
                    \multicolumn{5}{c}{Wilson Kernel} \\
                    \cmidrule(r){1-5}
                    $n$ & $1$ & $2$ & $3$ & $4$ \\
                    \midrule
                    PF & 15.4 & 21.6 & 24.8 & 26.8 \\
                    MU & 27.5 & $>297$ & $>6144$ & $>6144$ \\
                    SF & 18.6 & 226.0 & $>2048$ & $>6144$ \\
                    \toprule
                    \multicolumn{5}{c}{Brillouin Kernel} \\
                    \cmidrule(r){1-5}
                    $n$ & $1$ & $2$ & $3$ & $4$ \\
                    \midrule
                    PF & 8.04 & 15.0 & 20.2 & 23.6 \\
                    MU & 2.45 & 9.73 & 27.9 & 68.3 \\
                    SF & 9.01 & 36.8 & 107.0 & 260.0 \\
                    \bottomrule
                \end{tabular}
            \end{table}
    
            \begin{figure*}[!ht]
              \centering
              \begin{minipage}{0.49\textwidth}
                \includegraphics[width=\linewidth]{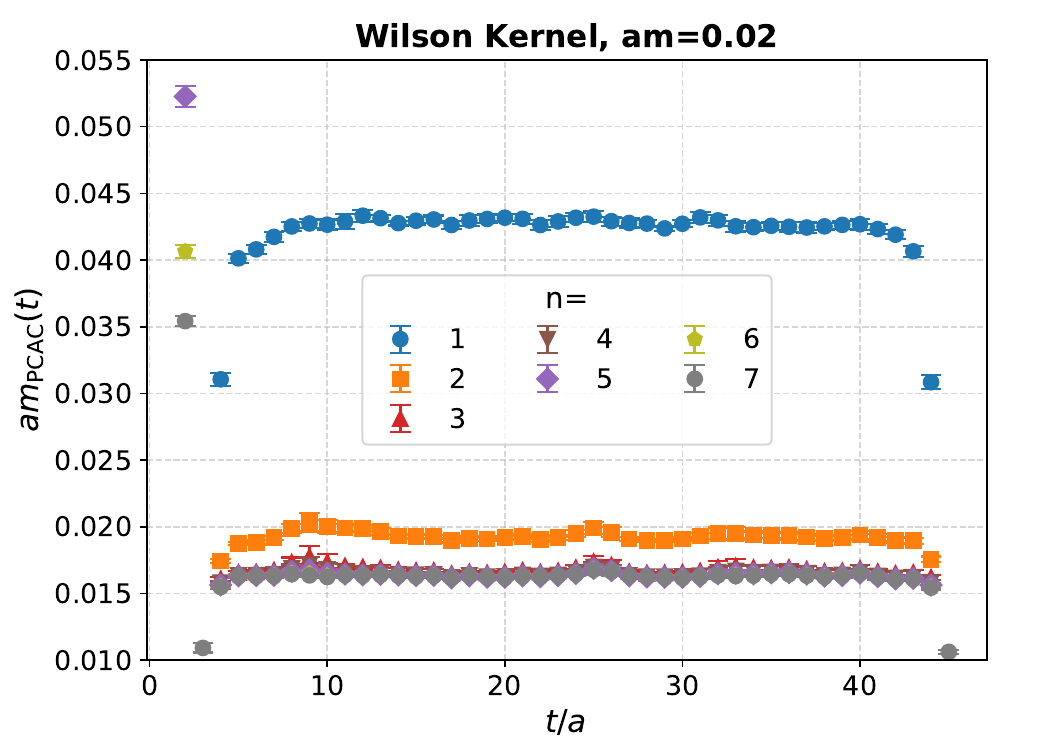}
              \end{minipage}
              \begin{minipage}{0.49\textwidth}
                \includegraphics[width=\linewidth]{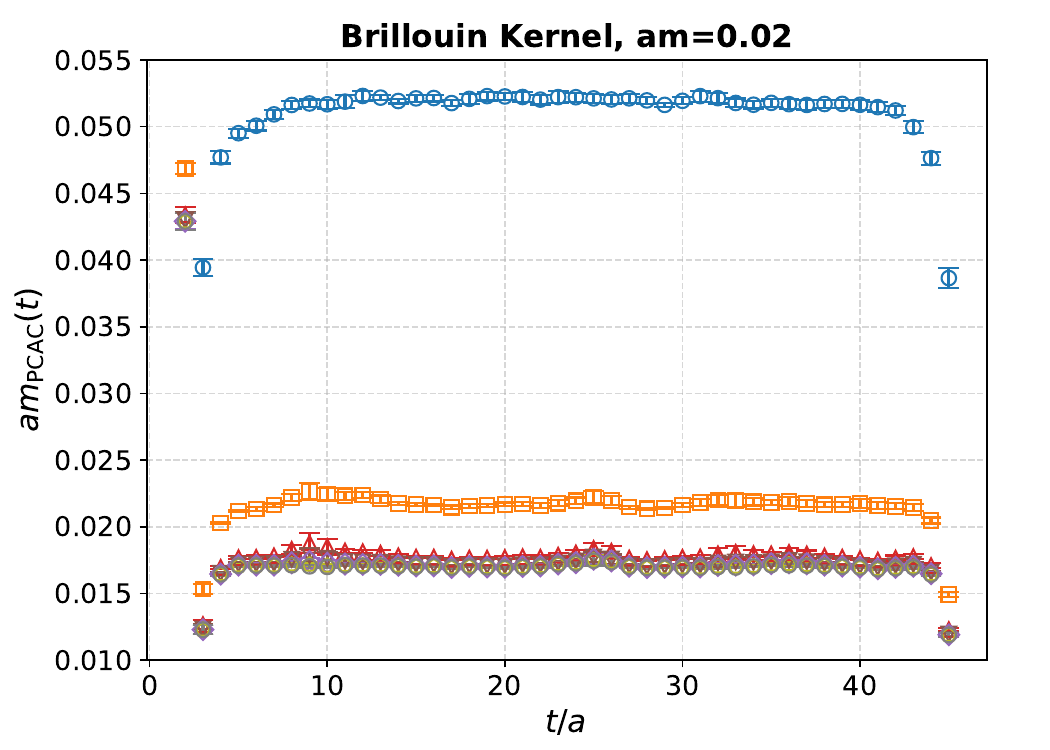}
              \end{minipage}
              
              \begin{minipage}{0.49\textwidth}
                \includegraphics[width=\linewidth]{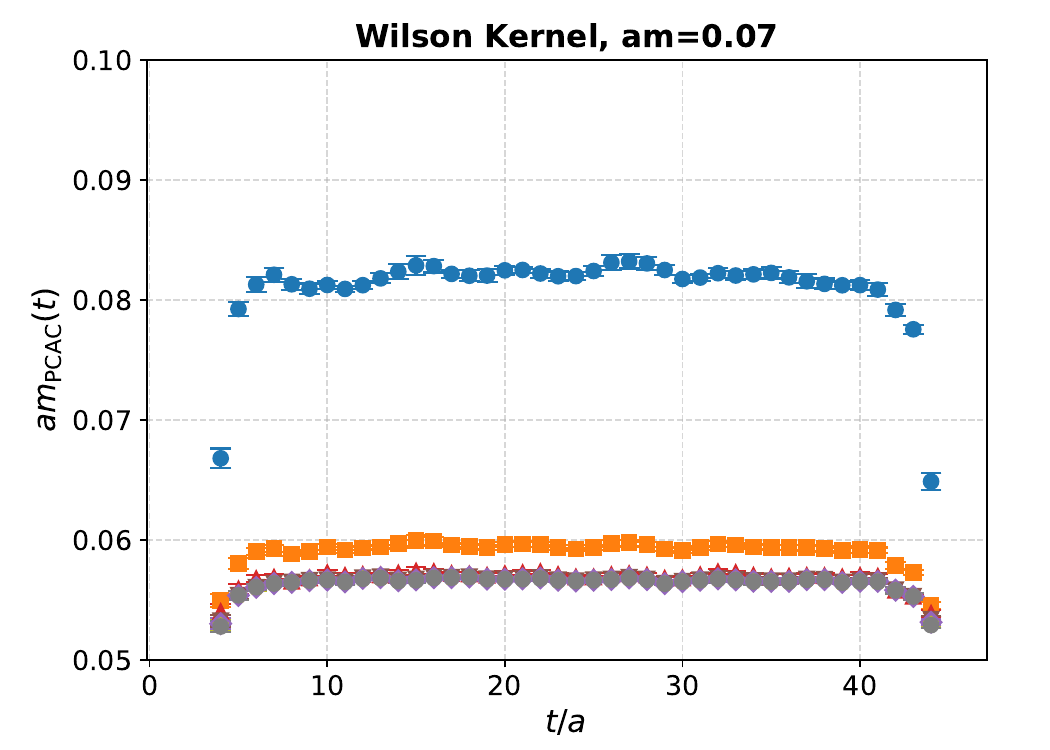}
              \end{minipage}
              \begin{minipage}{0.49\textwidth}
                \includegraphics[width=\linewidth]{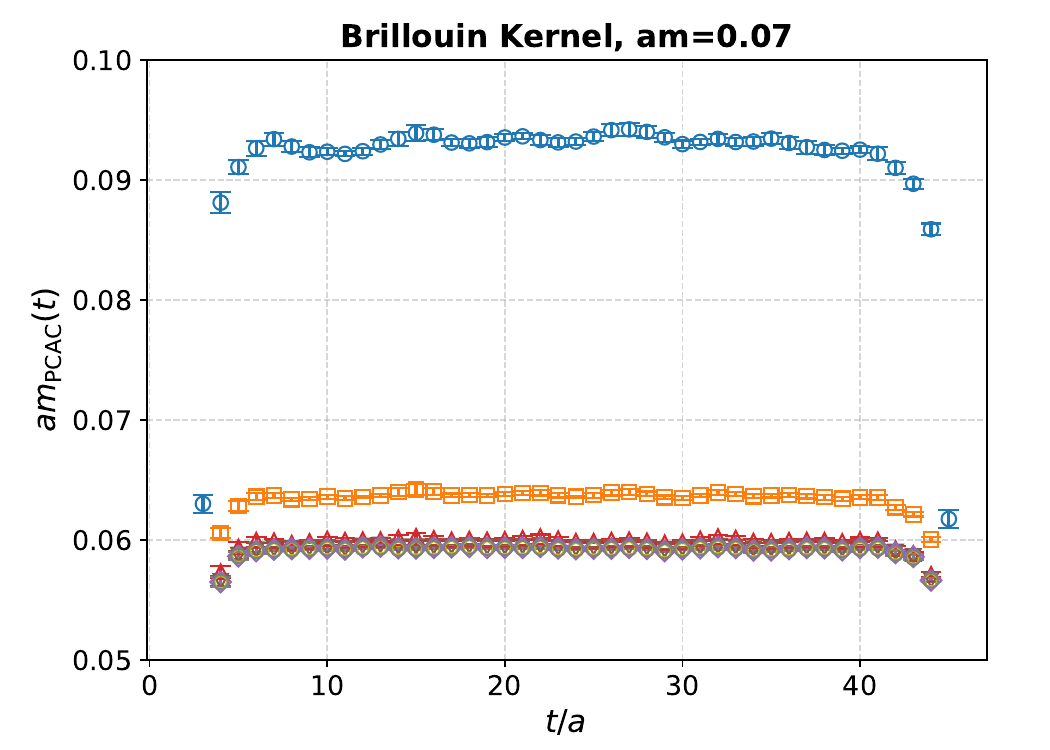}
              \end{minipage}
              \caption{PCAC mass $am_\mathrm{PCAC}(t)$ as defined in Eq.~\ref{eq:effective_PCAC_mass_definition} for the diagonal KL orders $n=1, \ldots, 7$ (as given in the legend) for the Wilson (left) and Brillouin (right) kernel for two values of the overlap bare quark mass, namely $am = 0.02$ (top) and $am = 0.07$ (bottom). Error bars indicate jackknife-estimated statistical uncertainties obtained over 25 configurations.}
              \label{fig:PCAC_mass_correlators_KL_overlap_n_and_mb}
            \end{figure*}
        
        \subsection{Inversion of the overlap operator}
            
            For solving $a D^{\mathrm{ov}}_m x = b$, a CGNR solver is employed (see Appendix~\ref{app:implementation}), in which every iteration requires an application of the overlap operator, with the matrix sign function approximated by diagonal KL iterates in partial fraction form in combination with an MSCG solver. For completeness, we provide here a comparison of this approach with two alternative approaches: one in which the single-fraction (SF) matrix sign function of Eq.~\ref{eq:KL_sign_function_single_fraction} is used, and one in which the SF expression is rearranged by multiplying both sides of $a D^{\mathrm{ov}}_m x = b$ from the left by $Q_{nn}(\mathbb{X}^2) \gamma_5$:
            \begin{gather}
                Q_{nn}(\mathbb{X}^2) \gamma_5 \, a D^{\mathrm{ov}}_m x \equiv M x = Q_{nn}(\mathbb{X}^2) \gamma_5 \, b \; ,
                \label{eq:multiply_up_trick}
            \end{gather}
            with:
            \begin{gather}
                M \equiv \left[ \left( \rho + \frac{a m}{2} \right) Q_{nn}(\mathbb{X}^2) \gamma_5 + \left(\rho - \frac{a m}{2}\right) \mathbb{X} \, P_{nn}(\mathbb{X}^2) \right] \; . \nonumber
            \end{gather}
            We refer to this latter approach as the ``multiply-up trick'' (MU). The three approaches are compared for a bare overlap quark mass of $am=0.15$ in Table~\ref{tab:partial_Vs_single_fraction_invert}. This analysis shows the advantage of the PF expansion over both alternative methods, with uncertainties not exceeding 5\%. More details on the inversion techniques employed are provided in Appendix~\ref{app:implementation}.
    
        \subsection{PCAC and pseudoscalar masses}
    
            We use the definition of $a m_{\mathrm{PCAC}}(t)$ for $c_\mathrm{SW}=0$:
            \begin{align}
                a m_{\mathrm{PCAC}}(t) = \frac{a}{2} \frac{\sum_{\vec{x}}\left\langle\bar{\partial}_4 A_4(\vec{x},t) P(0)\right\rangle}{\sum_{\vec{x}}\langle P(\vec{x},t) P(0)\rangle}
                \label{eq:effective_PCAC_mass_definition}
            \end{align}
            where $\bar{\partial}$ is discretized using the $\mathcal{O}(a^4)$ centered finite difference approximation:
            \begin{align}
                \bar{\partial} f(t) = \frac{1}{12} &\left[ f(t-2) - 8 f(t-1) \right. \nonumber\\
                &\quad\quad\quad\left. + 8 f(t+1) - f(t+2) \right] \; .
                \label{eq:symmetric_quartic_derivative}
            \end{align}
            For the pion effective mass, we use:
            \begin{align}
                &aM^\mathrm{eff}_{\pi}(t) = \nonumber \\                %
                &\quad\frac{1}{2}\log\!\left(  \frac{C(t-1)+\sqrt{C^2(t-1)-C^2(T/2)}}{C(t+1)+\sqrt{C^2(t+1)-C^2(T/2)}} \right) \; ,
                \label{eq:pion_effective_mass_definition}
            \end{align}
            with the shorthand $C(t) \equiv \sum_{\vec{x}}\langle P(\vec{x},t) P(0)\rangle$ for the two-point correlation function.
    
            \begin{figure*}[ht]
              \centering
              \begin{minipage}{0.49\textwidth}
                \centering
                \includegraphics[width=\linewidth]{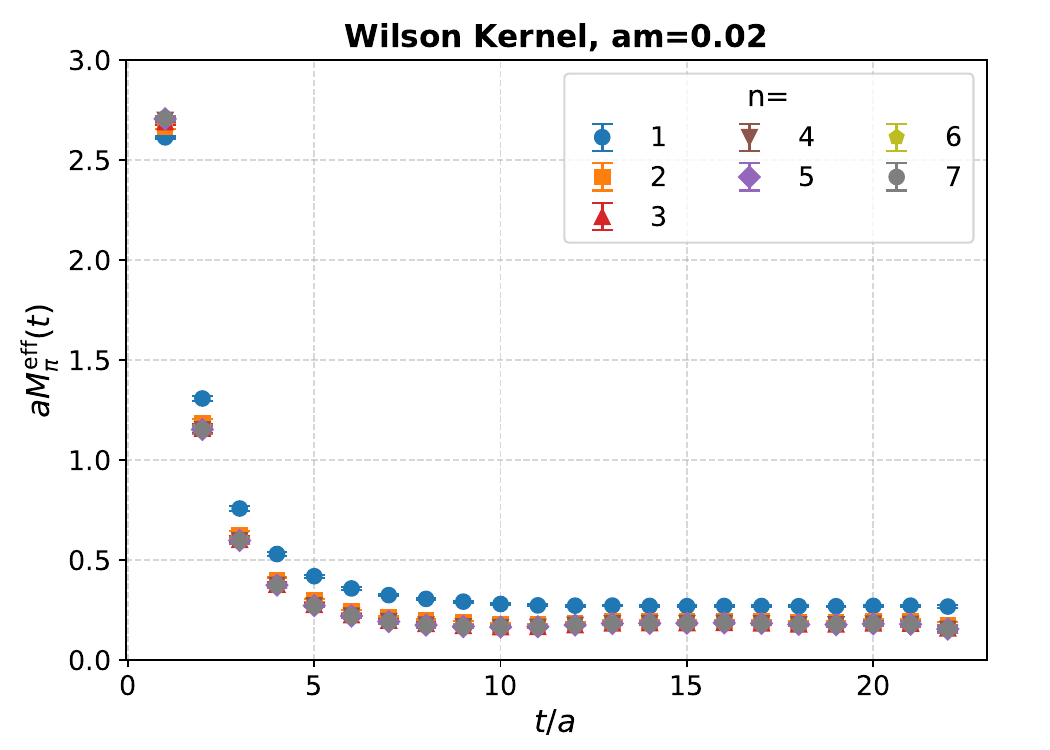}
              \end{minipage}
              \begin{minipage}{0.49\textwidth}
                \centering
                \includegraphics[width=\linewidth]{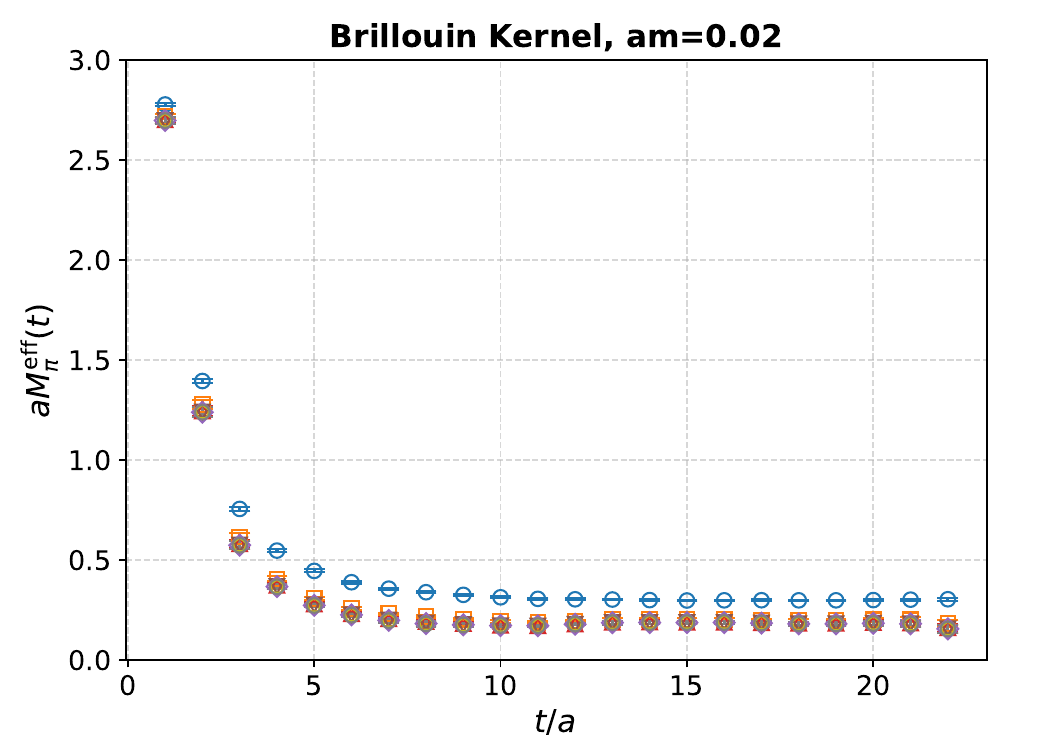}
              \end{minipage}
            
              \caption{Pion effective mass $aM^\mathrm{eff}_{\pi}(t)$, as defined in Eq.~\ref{eq:pion_effective_mass_definition}, for the diagonal KL orders $n=1, \ldots, 6$ for Wilson (left) and Brillouin (right) kernels at overlap bare quark mass $am = 0.02$. Error bars represent jackknife-estimated statistical uncertainties.}
              \label{fig:pion_effective_mass_correlators}
            \end{figure*}
    
            \begin{figure*}[ht]
              \centering
              \begin{minipage}{0.49\textwidth}
                \centering
                \includegraphics[width=\linewidth]{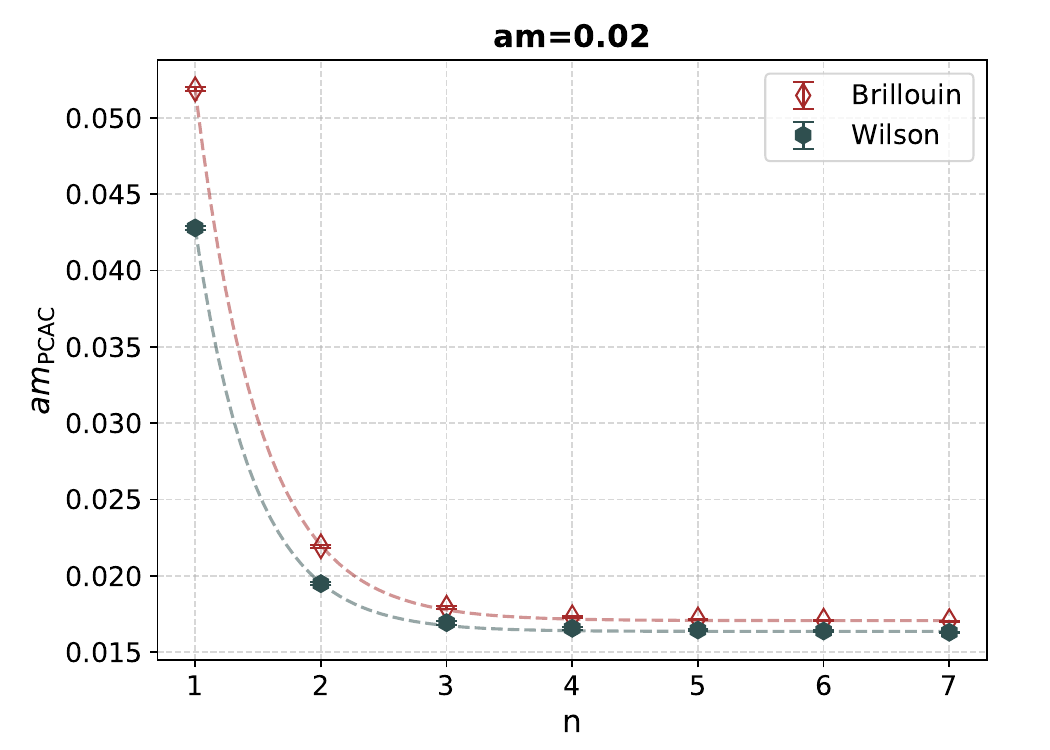}
              \end{minipage}
              \begin{minipage}{0.49\textwidth}
                \centering
                \includegraphics[width=\linewidth]{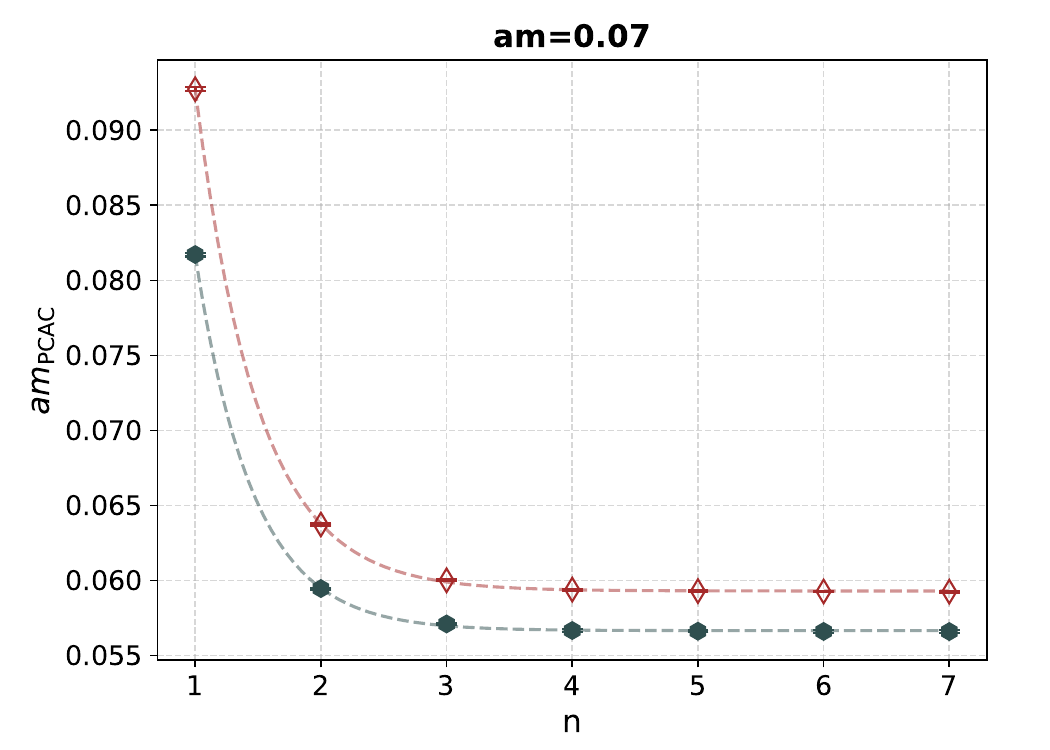}
              \end{minipage}
              
              \caption{PCAC mass values, extracted from plateau fits to the $am_\mathrm{PCAC}(t)$ data shown in Fig.~\ref{fig:PCAC_mass_correlators_KL_overlap_n_and_mb}, versus the diagonal KL order $n$ for overlap bare mass values $am = 0.02$ (left) and $am = 0.07$ (right), with open/closed symbols for the Brillouin/Wilson kernel. Cubic spline interpolations are shown to guide the eye. Error bars represent jackknife-estimated statistical uncertainties.}
              \label{fig:PCAC_mass_Vs_KL_diagonal_order}
            \end{figure*}
            
            In Fig.~\ref{fig:PCAC_mass_correlators_KL_overlap_n_and_mb} we show representative examples of the time-dependence of the PCAC mass ($am_{\mathrm{PCAC}}(t)$) for two values of the overlap bare quark mass, namely $am=0.02$ and $am=0.07$, and for diagonal KL orders $n=1$ to $n=7$. The errors shown are jackknife, obtained by inverting the overlap operator for 12 sources on 25 configurations for each kernel, value of $am$, and $n$.
    
            For both kernels, we see a clear plateau region for all values of $n$ between sink-source separations of about $t/a\in[10, T/a-10]$, where $T/a=48$ is the lattice time extent. Furthermore, we see that for $n=3$ and greater, the value of $a m_{\mathrm{PCAC}}(t)$ in the plateau region is approximately constant, an indication that the residual PCAC mass at these approximation orders is of the same order as the statistical errors. A systematic study of the PCAC mass as a function of the overlap bare quark mass $am$ is provided further on.
    
            The effective pion mass ($aM^\mathrm{eff}_{\pi}(t)$) is shown in Fig.~\ref{fig:pion_effective_mass_correlators} at overlap bare quark mass $am=0.02$ using both kernels, where in this case we symmetrize the correlator over the mid-point of the time extent. As can be seen, the curves for the effective pion mass coincide after $n=3$, suggesting that the dependence of the pion meson mass on $n$ follows that of $am_\mathrm{PCAC}$. Indicatively, in physical units, the effective mass plateaus at around $m_\pi \simeq 600$~MeV for $n\ge 3$ and $am=0.02$.
    
            We determine the PCAC mass $am_\mathrm{PCAC}$ by fitting to the plateau region of the (time-dependent) PCAC mass $a m_{\mathrm{PCAC}}(t)$ of Fig.~\ref{fig:PCAC_mass_correlators_KL_overlap_n_and_mb} for the available $n=1,\ldots,7$. The values obtained are plotted in Fig.~\ref{fig:PCAC_mass_Vs_KL_diagonal_order} as a function of $n$ for the two values of the overlap bare quark masses used in Fig.~\ref{fig:PCAC_mass_correlators_KL_overlap_n_and_mb}. As can be seen, $am_\mathrm{PCAC}$ approaches its asymptotic value exponentially. Furthermore, for a given $am$, the asymptotic value is different between Wilson and Brillouin, indicating that the slope of $am_\mathrm{PCAC}$ versus $am$ is different for the two kernels.
    
\begin{figure*}[ht]
                \centering
                \begin{minipage}{0.49\textwidth}
                    \centering
                    \includegraphics[width=\linewidth]{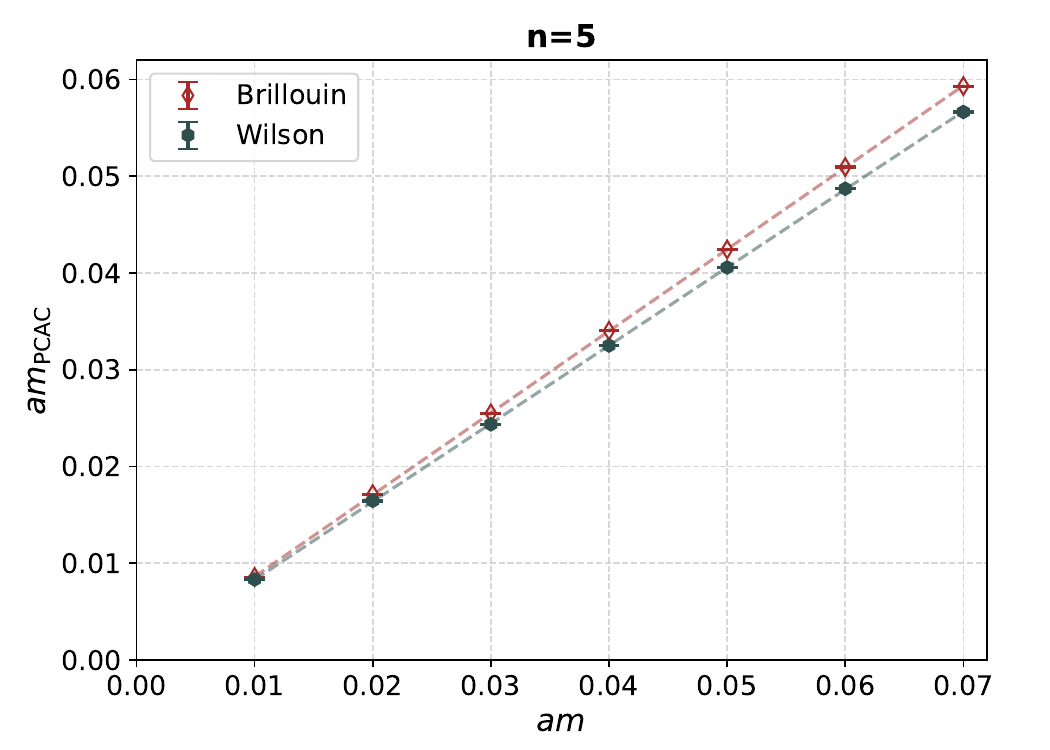}
                \end{minipage}
                \begin{minipage}{0.49\textwidth}
                    \centering
                    \includegraphics[width=\linewidth]{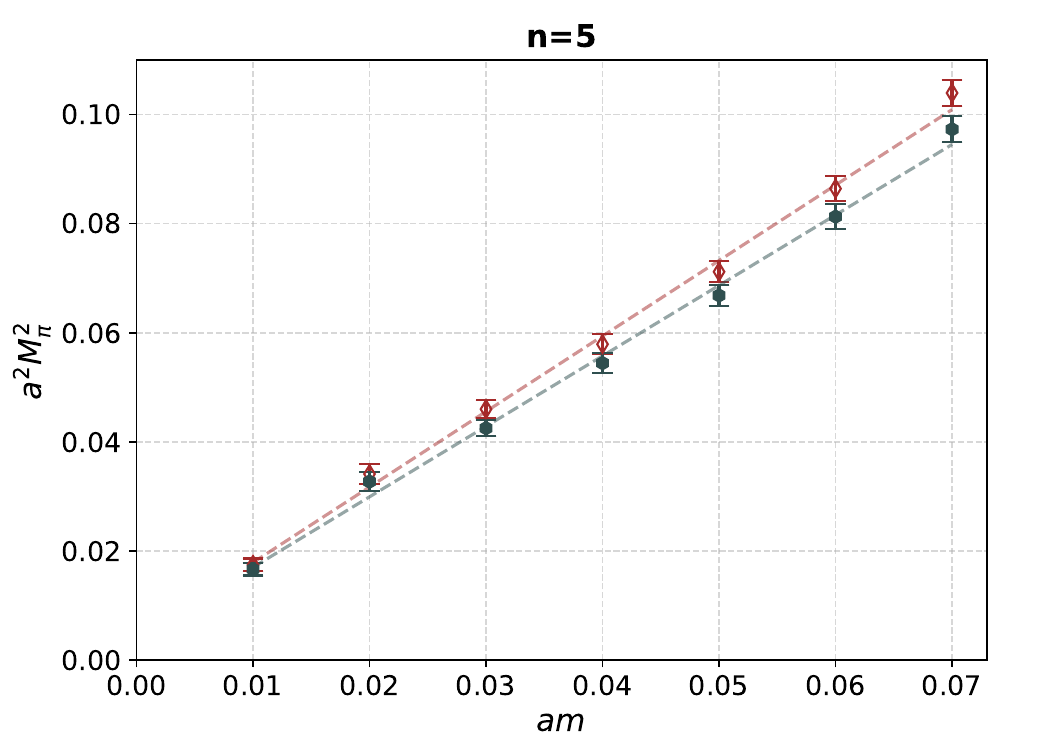}
                \end{minipage}
                
                \caption{PCAC mass (left panel) and pion mass squared (right panel) plotted against the overlap bare mass $am$ for the Wilson kernel (filled markers) and the Brillouin kernel (empty markers) at fixed diagonal KL order $n = 5$. Straight dashed lines are meant to guide the eye. Error bars represent jackknife-estimated uncertainties. The legend applies to both panels.}
                \label{fig:pcac_and_pion_Vs_bare_mass}
            \end{figure*}

            This kernel-dependent behavior becomes more evident in Fig.~\ref{fig:pcac_and_pion_Vs_bare_mass}, which shows that as the overlap bare mass approaches zero, both the PCAC mass (left panel) and pion mass squared (right panel) values for different kernels converge, suggesting that kernel effects diminish in the chiral limit.
            
            \begin{figure*}[ht]
                \centering
                \begin{minipage}{0.49\textwidth}
                    \centering
                    \includegraphics[width=\linewidth]{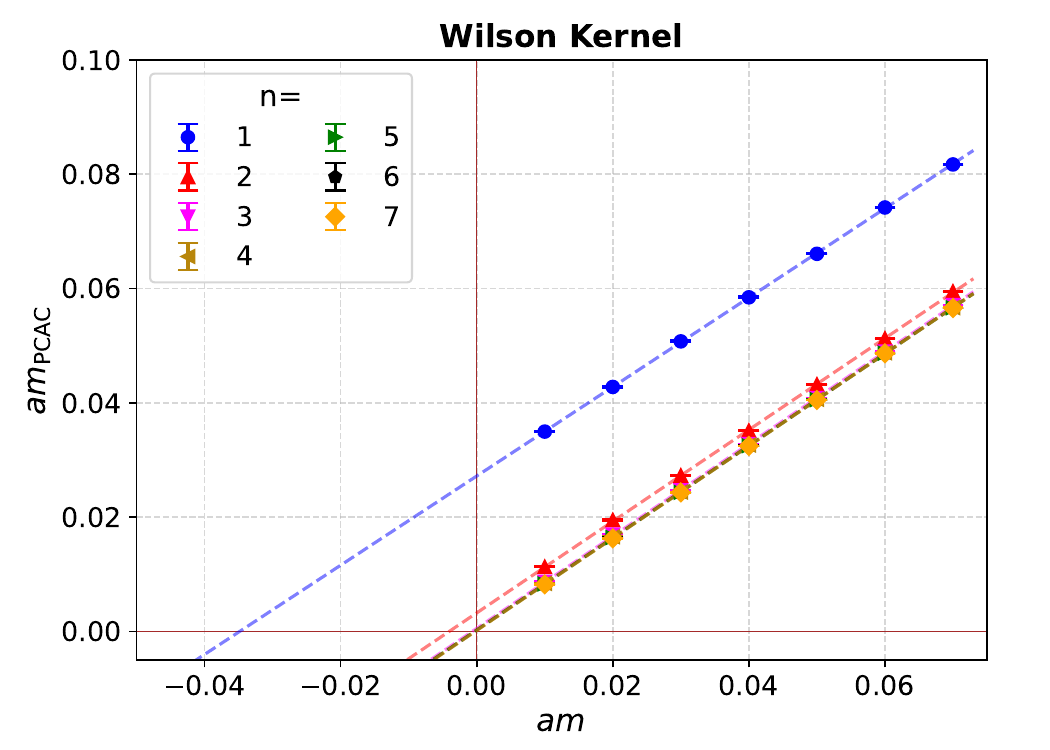}
                \end{minipage}
                \begin{minipage}{0.49\textwidth}
                    \centering
                    \includegraphics[width=\linewidth]{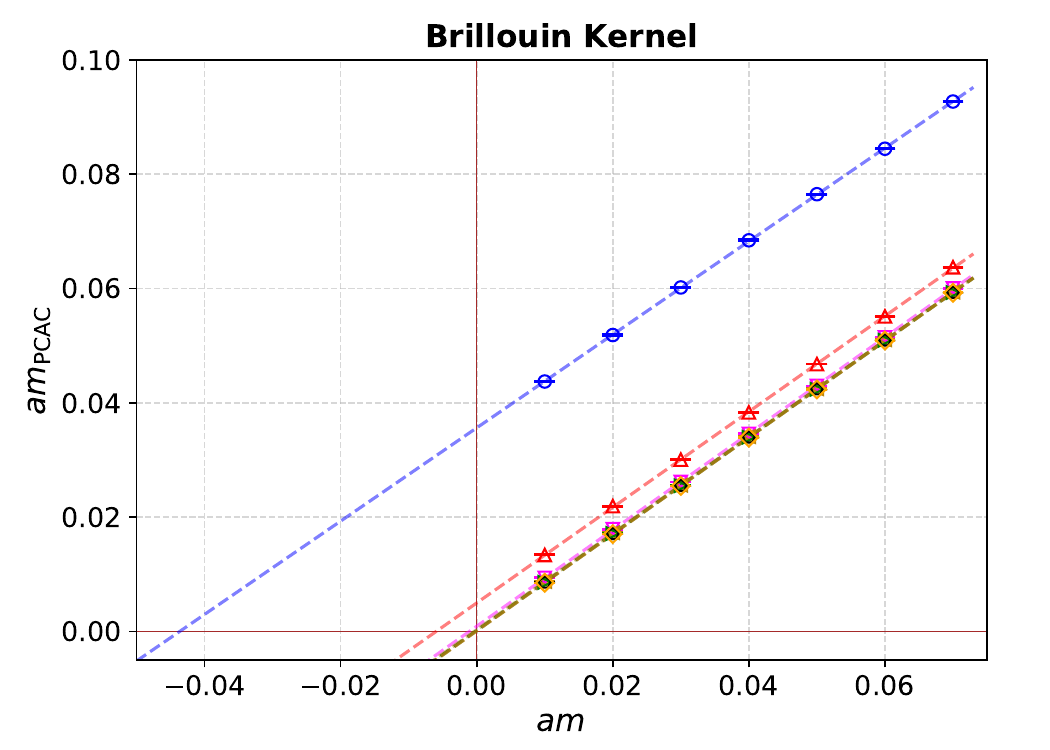}
                \end{minipage}
            
                \caption{PCAC mass values plotted against the overlap bare mass, for the Wilson (left panel) and Brillouin (right panel) kernel, across the diagonal KL orders $n=1, \ldots, 7$. Straight dashed lines represent linear fits to the data. The point on the x-axis where the y-value is zero is referred to as the critical bare mass. Error bars represent jackknife-estimated statistical uncertainties.}
                \label{fig:KL_pcac_Vs_Bare_mass}
            \end{figure*}
    
            Fig.~\ref{fig:KL_pcac_Vs_Bare_mass} examines this relationship across different diagonal KL orders, separately for each kernel, revealing that while the slope of these curves remains consistent with increasing $n$, the intercept decreases systematically in absolute value. This behavior enables reliable extrapolation to determine $am_{\mathrm{bare}}^{\text{critical}}$, the critical approximate-overlap bare mass that corresponds to $a m_{\mathrm{PCAC}} = 0$. Consequently, the residual PCAC mass can be defined as the y-intercept $am_{\mathrm{PCAC}}$ for $am=0$. For each bare mass value, a different set of gauge field configurations is used, allowing for uncorrelated linear fits.
            
            \begin{figure*}[ht]
                \centering
                \begin{minipage}{0.49\textwidth}
                    \centering
                    \includegraphics[width=\linewidth]{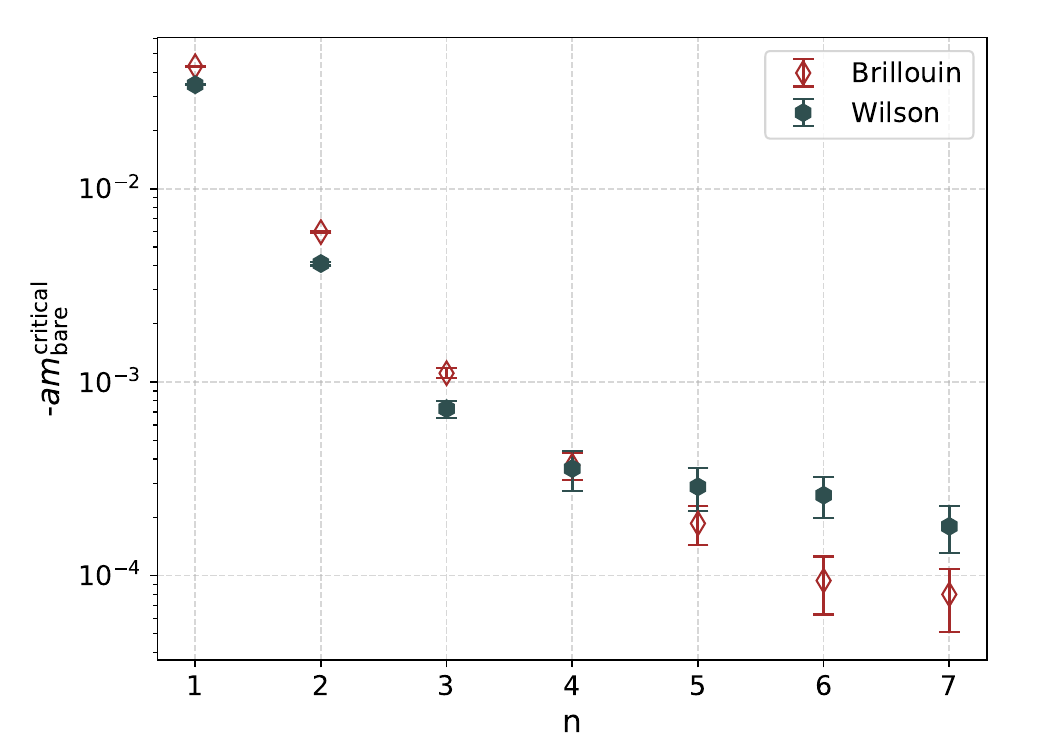}
                \end{minipage}
                \begin{minipage}{0.49\textwidth}
                    \centering
                    \includegraphics[width=\linewidth]{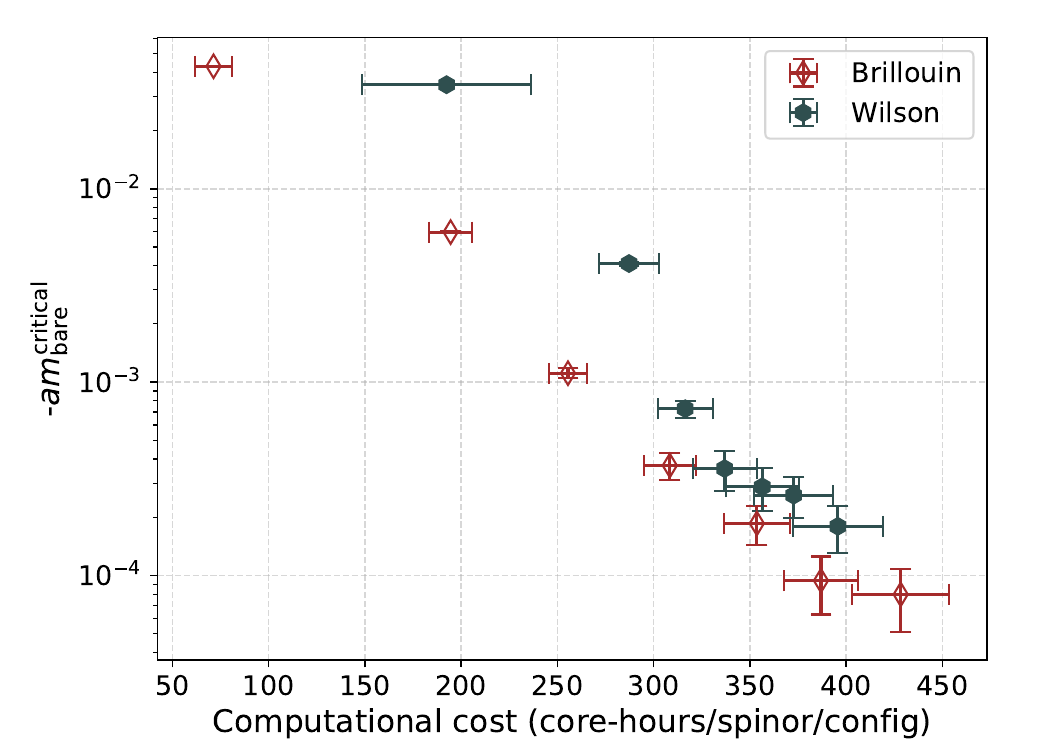}
                \end{minipage}
            
                \caption{Left: Negative critical bare masses ($-am^{\mathrm{critical}}_{\mathrm{bare}}$), extracted from the data in Fig.~\ref{fig:KL_pcac_Vs_Bare_mass}, as a function of the diagonal KL order $n$ for the Brillouin kernel (open symbols) and the Wilson kernel (full symbols). Right: the same data plotted against the computational cost (in core-hours per spinor per configuration), as defined in the text.}
                \label{fig:critical_bare_mass_Vs_KL_diagonal_order}
            \end{figure*}
    
            With the value of the critical bare mass for each value of $n$ at hand, we can investigate how $am^{\mathrm{critical}}_{\mathrm{bare}}$ approaches zero as $n$ is increased. This is shown in Fig.~\ref{fig:critical_bare_mass_Vs_KL_diagonal_order} (left panel), where we also associate a computational cost for each value of $n$ (right panel). For the purposes of the latter, since determining the critical bare mass requires PCAC mass calculations at multiple bare mass values, each with different computational costs, we define a consistent cost metric for each value of $n$. Namely, we measure the core-hours required to calculate the PCAC mass at a given bare mass that corresponds to a pion mass of approximately $300$~MeV. This metric provides for a physically motivated definition of the computational cost, namely for a given pion mass. The horizontal errors in the right panel of Fig.~\ref{fig:critical_bare_mass_Vs_KL_diagonal_order} are obtained when propagating the error for the bare mass that yields the target pion mass.
    
            \begin{figure*}[ht]
                \centering
                \begin{minipage}{0.49\textwidth}
                    \centering
                    \includegraphics[width=\linewidth]{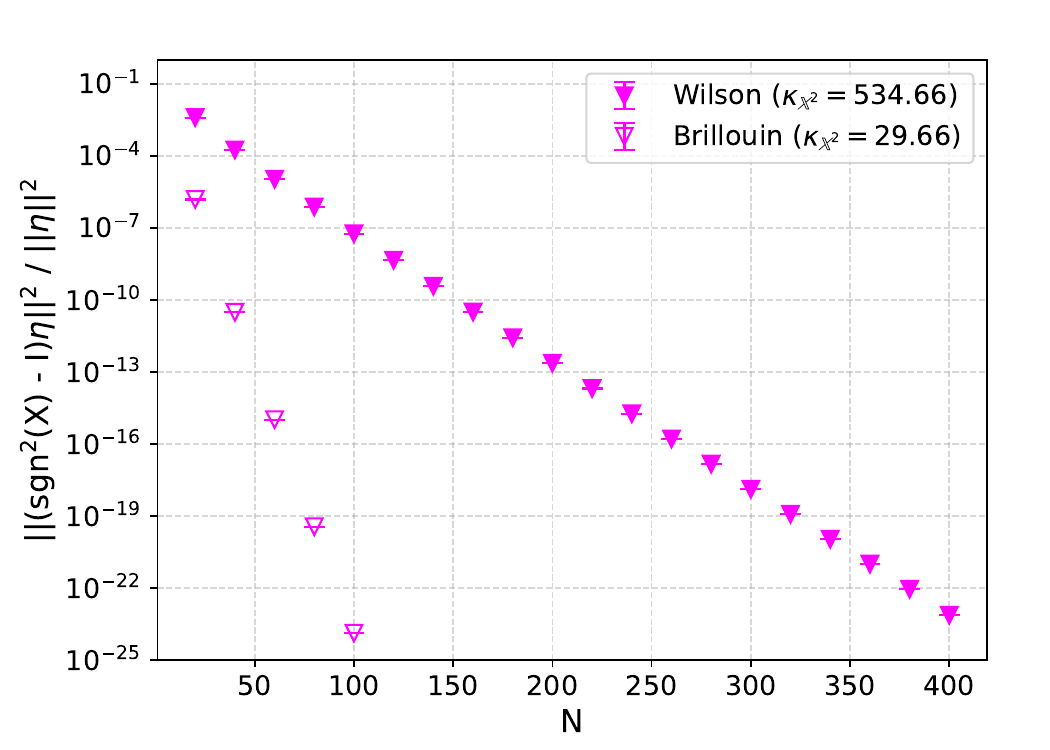}
                \end{minipage}
                \hfill
                \begin{minipage}{0.49\textwidth}
                    \centering
                    \includegraphics[width=\linewidth]{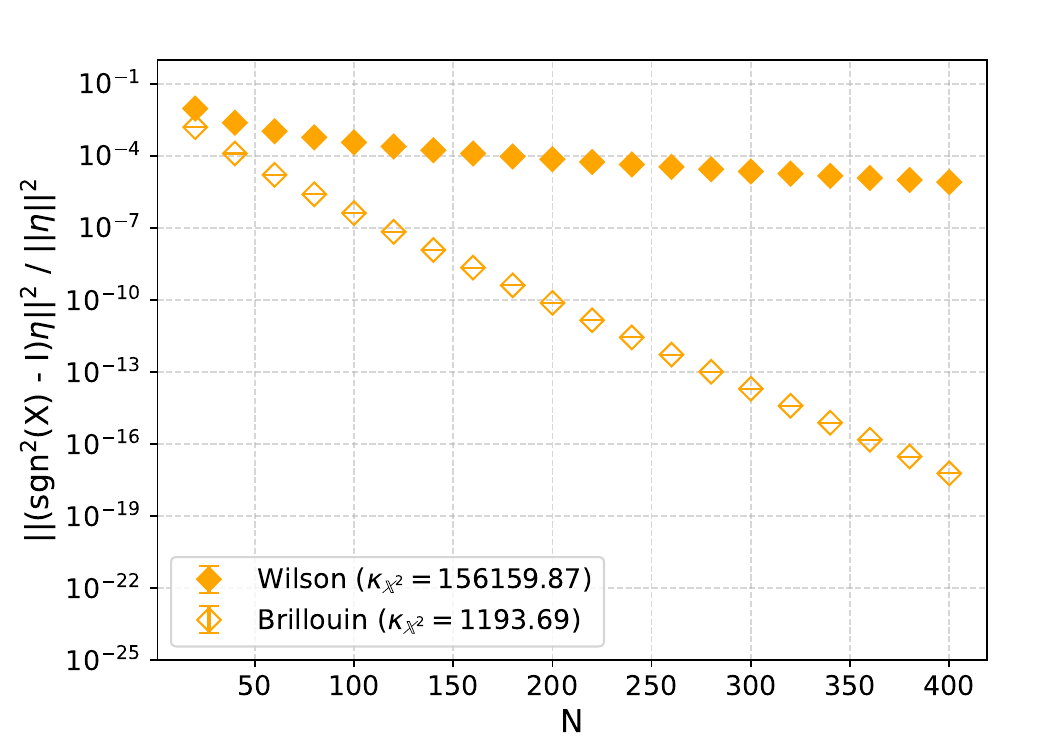}
                \end{minipage}
    
                \caption{Same as Fig.~\ref{fig:KL_sign_squared_Vs_n_by_kernel}, but for Chebyshev polynomial expansions of order $N$ rather than diagonal KL approximations of order $n$. Results are shown for the same gauge field configurations as in Fig.~\ref{fig:KL_sign_squared_Vs_n_by_kernel}, with matching colors and marker styles for comparison.}
                \label{fig:Chebyshev_sign_squared_Vs_number_of_terms_by_kernel}
            \end{figure*}
    
            Both panels show a fast and monotonic improvement with increasing diagonal KL order, demonstrating systematic enhancement of the approximation quality. While critical bare mass values do not differ significantly between kernels at the same diagonal KL order, we find that the Brillouin kernel provides a non-negligible computational advantage, particularly for orders $n \leq 4$. As an indication, we take $n=4$ as a reference approximation order, since by this order physical observables have effectively converged toward their asymptotic values (cf.~Fig.~\ref{fig:PCAC_mass_Vs_KL_diagonal_order}); both kernels reach $|am_{\mathrm{bare}}^{\mathrm{critical}}| \approx 4\times10^{-4}$ at this order, with the Brillouin kernel requiring approximately $310$ core-hours per spinor per configuration compared to about $340$ for the Wilson kernel; a reduction of roughly $9$\%. This precision level will also serve as the reference for the cross-method comparison in Sec.~\ref{subsec:combined_physical_observables}.
    
            These results provide evidence that the Brillouin kernel enables the construction of a more chiral overlap operator with reduced computational cost compared to the Wilson kernel, at least within the approximation orders investigated.
    
    
    \section{Comparison with Chebyshev Polynomial Approximation}
    \label{sec:comparison}
    
        Having established the feasibility of using diagonal KL iterates as an approximation to the sign function, we proceed to compare it to a well-established method in the literature, namely the Chebyshev polynomial approximation. A brief review of the Chebyshev method is provided in Appendix~\ref{app:Chebyshev}. For the purposes of our comparison to the KL approximation, we will investigate the properties of the Chebyshev approximation as a function of the order of the Chebyshev polynomial expansion ($N$). For the definition, see Eq.~\ref{eq:Chebyshev_sign_function_approximation} in Appendix~\ref{app:Chebyshev}. We remind the reader that approximating the sign function with a Chebyshev polynomial expansion of order $N$ requires applying $\mathbb{X}^k$ for all odd powers $k$ from $k=1$ to $k=2N-1$.
    
            \begin{figure*}[ht]
                \centering
                \begin{minipage}{0.49\textwidth}
                    \centering
                    \includegraphics[width=\linewidth]{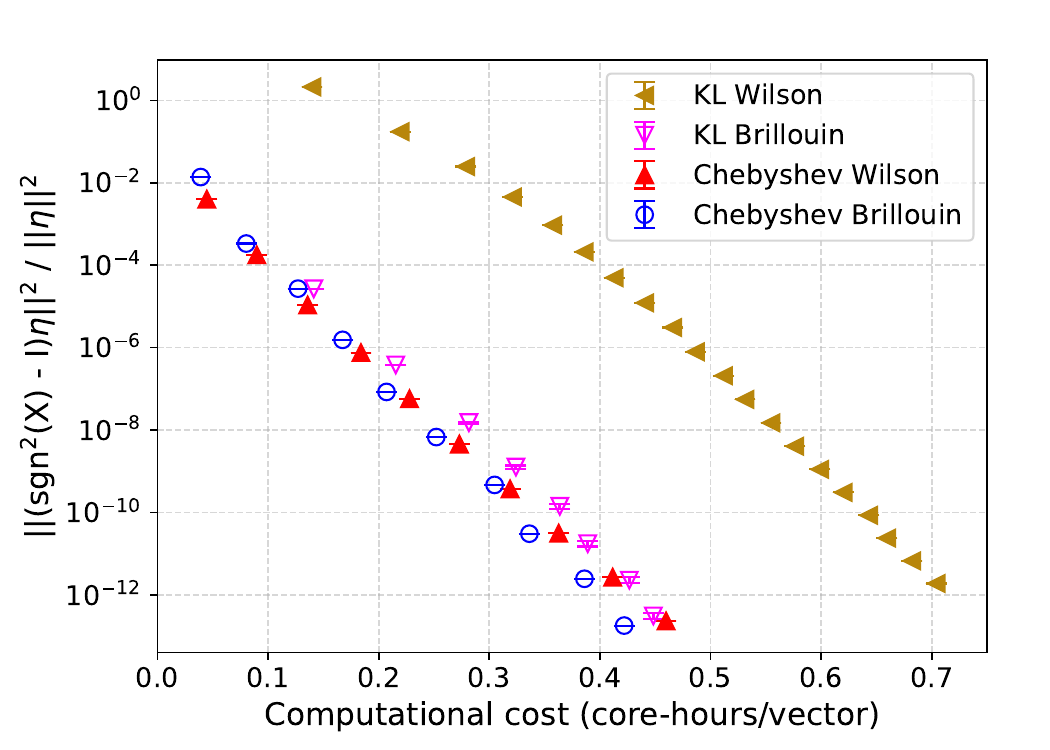}
                \end{minipage}
                \begin{minipage}{0.49\textwidth}
                    \centering
                    \includegraphics[width=\linewidth]{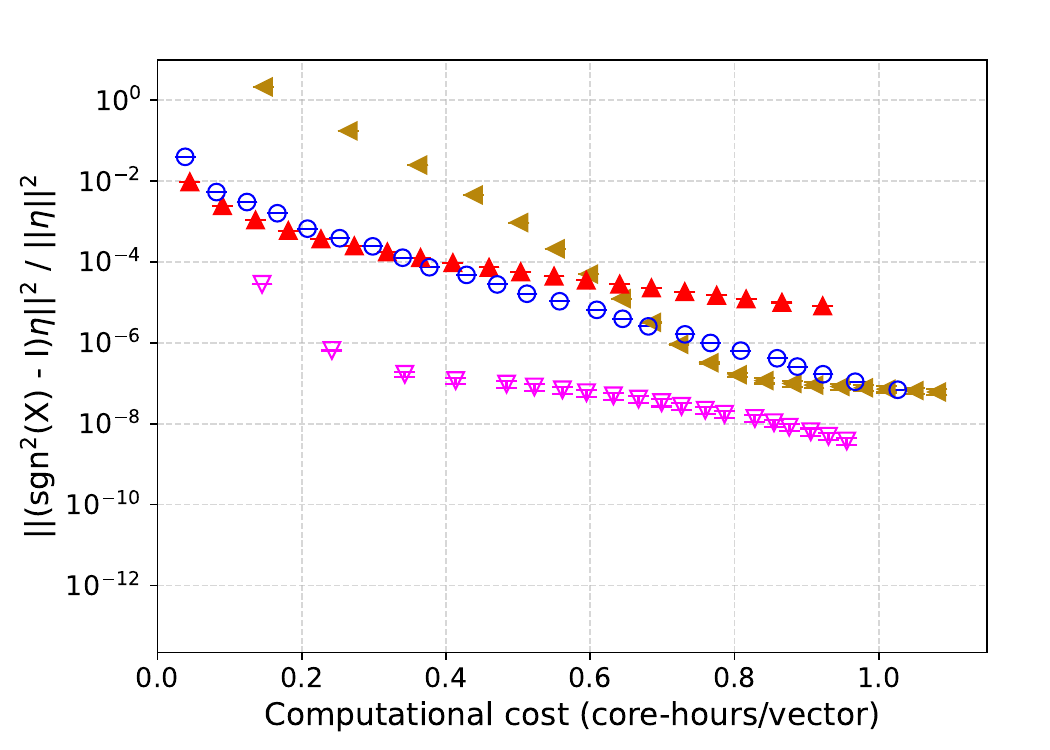}
                \end{minipage}
            
                \caption{Same as Fig.~\ref{fig:KL_sign_squared_Vs_core_hours}, but augmented by the Chebyshev data. Results for two configurations (a well-conditioned one on the left, a worse-conditioned one on the right) and the four possible combinations of Wilson or Brillouin kernel and Kenney-Laub (KL) or Chebyshev approximation are shown (as indicated by the legend). For KL the data represent the diagonal KL orders $n=1,2,\ldots,20$, while for Chebyshev the data represent the polynomial orders $N=10, 20,\dots, 400$. For the Chebyshev method, the plotted cost excludes the overhead of the Lanczos eigenvalue estimation (see Appendix~\ref{app:implementation}), which amounts to 0.71 (Wilson) and 0.47 (Brillouin) core-hours for the left panel, and 0.70 (Wilson) and 0.60 (Brillouin) core-hours for the right panel. The KL method incurs negligible overhead.}
                \label{fig:Combined_sign_squared_Vs_cost}
            \end{figure*}
            
            \begin{figure*}[ht]
                \centering
                \begin{minipage}{0.49\textwidth}
                    \centering
                    \includegraphics[width=\linewidth]{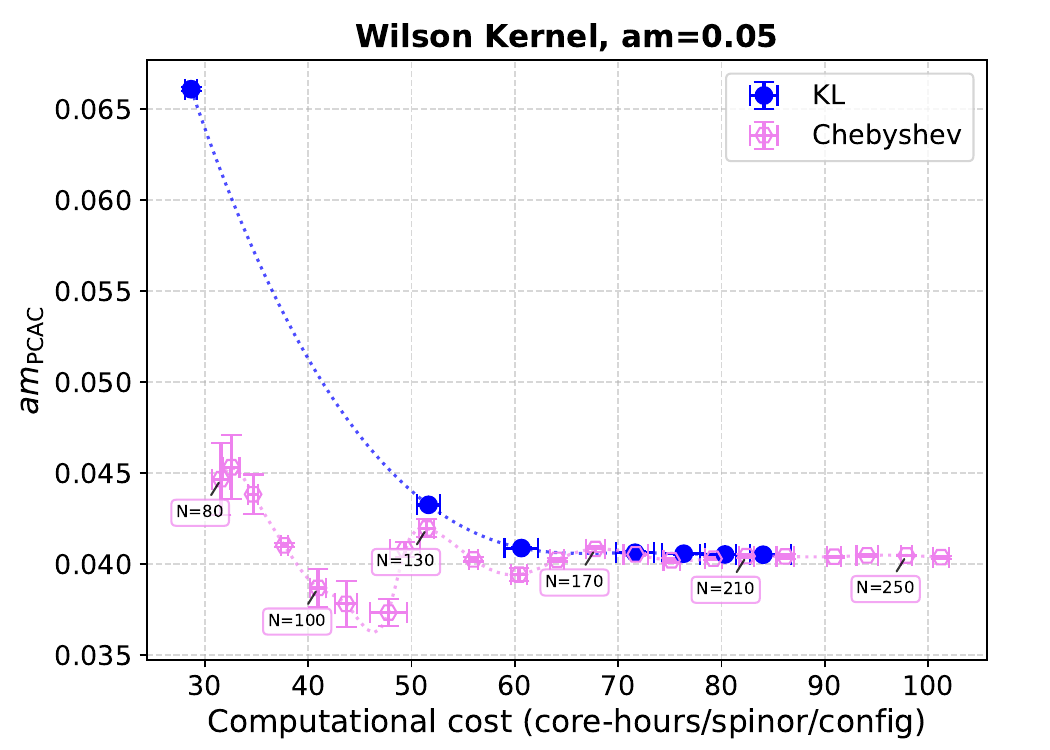}
                \end{minipage}
                \begin{minipage}{0.49\textwidth}
                    \centering
                    \includegraphics[width=\linewidth]{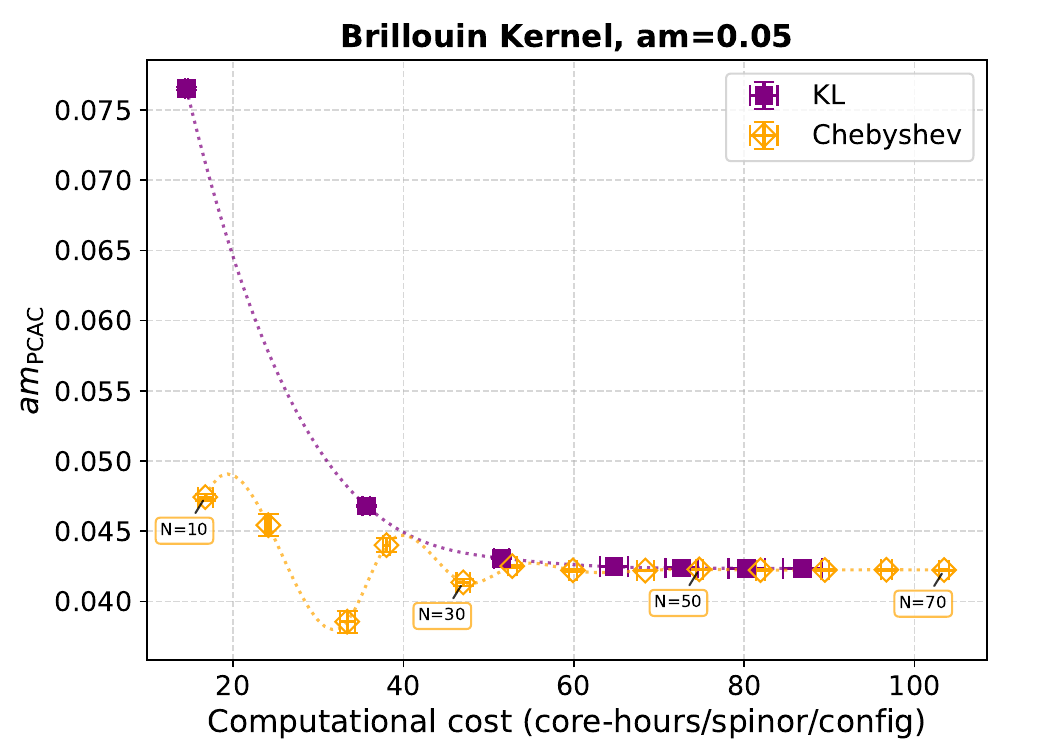}
                \end{minipage}
            
                \caption{PCAC mass values plotted against computational cost (in core-hours per spinor per configuration) at the overlap bare mass $am = 0.05$ for Wilson kernel (left) and Brillouin kernel (right). Each panel compares diagonal Kenney-Laub orders $n = 1, 2, \ldots, 7$ (filled markers) with Chebyshev orders $N$ as indicated in the annotation boxes (empty markers). Cubic splines are added to guide the eye. Vertical error bars are jackknife-estimated statistical uncertainties of the PCAC mass, while horizontal error bars are the error in the cost over multiple configurations and spinors.}
                \label{fig:Combined_PCAC_mass_Vs_cost}
            \end{figure*}
    
        \subsection{Ginsparg-Wilson relation and sign function convergence}       
    
            \begin{figure*}[ht]
                \centering
                \begin{minipage}{0.49\textwidth}
                    \centering
                    \includegraphics[width=\linewidth]{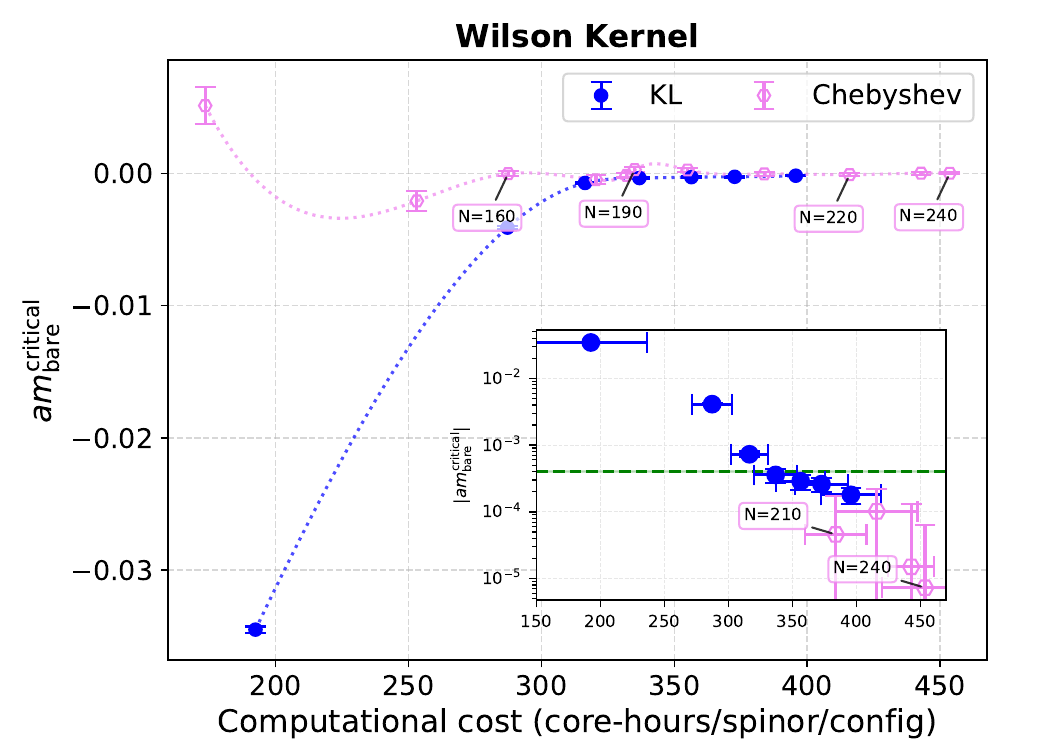}
                \end{minipage}
                \begin{minipage}{0.49\textwidth}
                    \centering
                    \includegraphics[width=\linewidth]{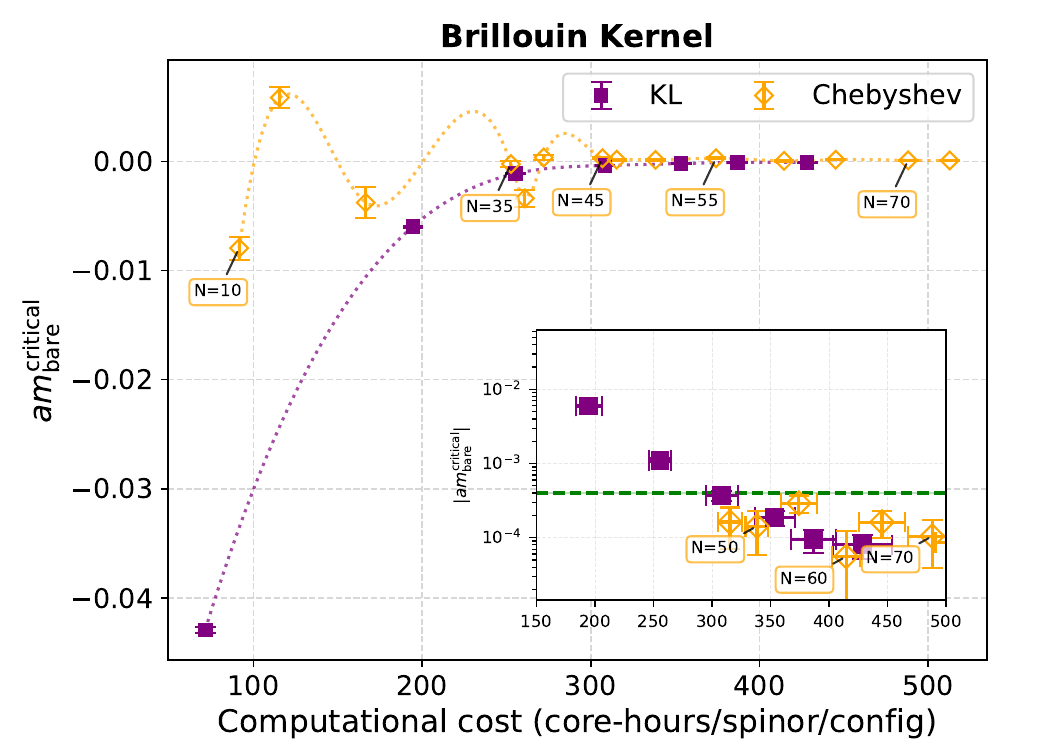}
                \end{minipage}
                
                \caption{Critical bare masses $am_{\mathrm{bare}}^{\mathrm{critical}}$, as extracted from the extrapolations shown in Fig.~\ref{fig:KL_pcac_Vs_Bare_mass}, plotted against the computational cost (in core-hours per spinor and configuration) for the Kenney-Laub (filled markers) and the Chebyshev (empty markers) approximations and the Wilson (left) and Brillouin (right) kernels. Either inset displays the absolute critical bare mass values on a semi-logarithmic scale for a selected range of computational cost values. The horizontal green line indicates the reference precision level to be used in Fig.~\ref{fig:absolute_critical_bare_mass_comparison_bar_plot}. Error bars represent properly propagated statistical uncertainties from the extrapolation procedure.}
                \label{fig:Combined_critical_bare_mass_Vs_cost}
            \end{figure*}
                
            In Fig.~\ref{fig:Chebyshev_sign_squared_Vs_number_of_terms_by_kernel} we compute the average sign-squared violation $\delta_{\mathrm{sgn}^2}$ over 10 random vectors per data point as a function of the Chebyshev polynomial order $N$ for the same gauge field configurations as those used in Fig.~\ref{fig:KL_sign_squared_Vs_n_by_kernel}. Consistent with our observations for the KL method, the Brillouin kernel is found to enable better convergence for a fixed Chebyshev polynomial order $N$ compared to the Wilson kernel.
                
            To facilitate a direct comparison between both approximation methods and kernels, we show in Fig.~\ref{fig:Combined_sign_squared_Vs_cost} the average sign-squared violation $\delta_{\mathrm{sgn}^2}$ as a function of the computational cost in core-hours per random vector. For the Chebyshev method, the plotted cost excludes the Lanczos eigenvalue estimation overhead, which is absent for the KL method. In general, the results indicate that when using the Brillouin kernel in either sign function approximation, a given $\delta_{\mathrm{sgn}^2}$ threshold is reached with fewer resources compared to when using the Wilson kernel. Comparing the KL and Chebyshev methods, the conclusion drawn from this plot is that the KL-Brillouin combination is more efficient for either configuration.
    
        \subsection{Physical observables and critical bare mass}
        \label{subsec:combined_physical_observables}
            
            We now directly compare the computational efficiency of both approximation methods for calculating physical observables. In Fig.~\ref{fig:Combined_PCAC_mass_Vs_cost} we show the computational cost (in core-hours per spinor per number of gauge field configurations) for calculating the PCAC mass at fixed bare mass $am=0.05$ using both approximation methods and kernels. The Brillouin kernel achieves earlier stabilization of PCAC mass values at a lower computational cost compared to the Wilson kernel, regardless of the approximation method. In addition, the KL approach demonstrates monotonic convergence for both kernels, in contrast to the Chebyshev method, which appears to oscillate towards the asymptotic PCAC mass value.
    
            We further examine the critical bare mass values $m_{\mathrm{bare}}^{\text{critical}}$ calculated for both methods for various values of the diagonal KL order $n$ and the Chebyshev polynomial order $N$. In Fig.~\ref{fig:Combined_critical_bare_mass_Vs_cost} we show the critical bare mass versus computational cost measured in core-hours per spinor per configuration, using the same computational cost definition as in the previous section (core-hours needed to calculate the PCAC mass at bare mass values corresponding to $\sim\!300$ MeV pion mass for each parameter value), enabling a direct comparison between the two methods and the two kernels.
    
            \begin{figure}[ht]
                \centering
                \includegraphics[width=\linewidth]{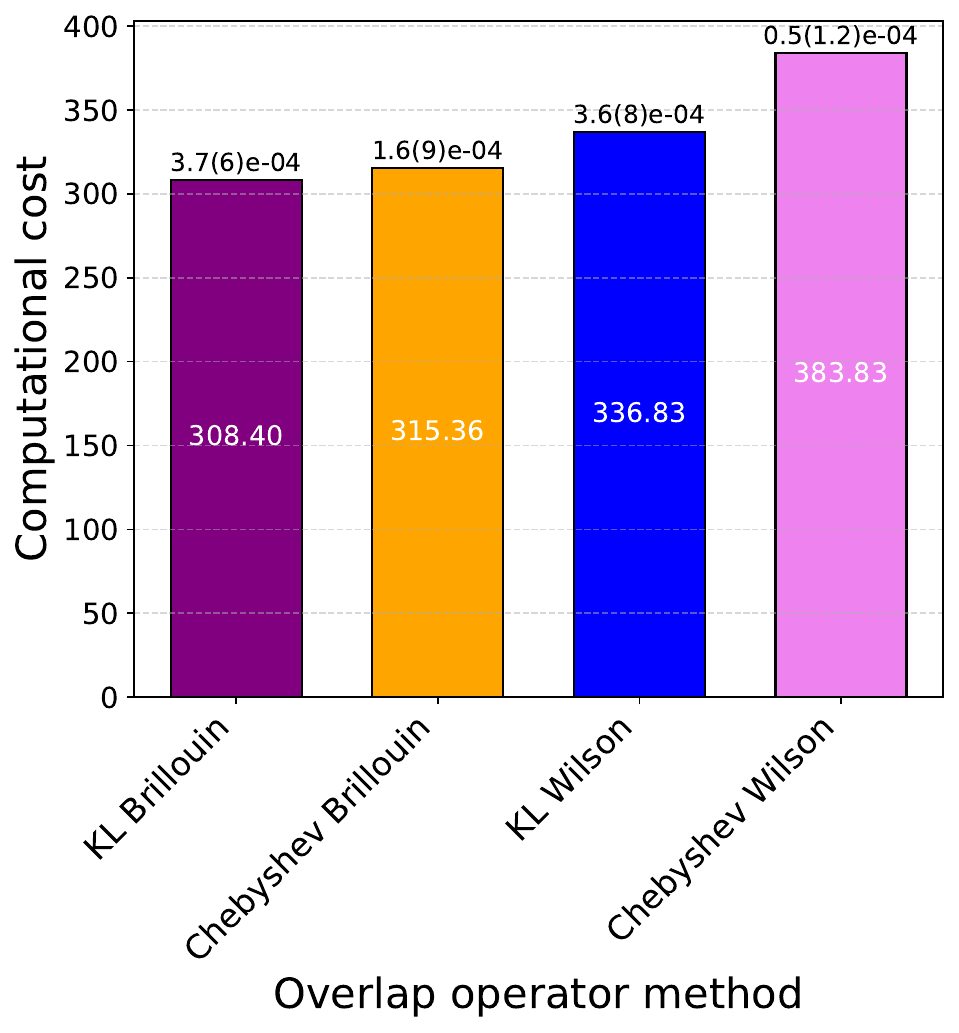}
    
                \caption{Bar plot of the computational cost (in core-hours per spinor and configuration) required to reach $|am_{\mathrm{bare}}^{\mathrm{critical}}| < 4\times10^{-4}$ for all four method-kernel combinations. The absolute critical bare value attained is indicated above each bar.}
                \label{fig:absolute_critical_bare_mass_comparison_bar_plot}
            \end{figure}
            
            As can be seen, the two methods are competitive in terms of the final precision obtained, once convergence to a critical bare mass value consistent with zero has been achieved. However, for the case of KL, convergence improves monotonically with increasing computational cost, in contrast to the oscillatory behavior of the Chebyshev method. The KL method thus yields a more predictable convergence toward the final critical bare mass value.
    
            Finally, Fig.~\ref{fig:absolute_critical_bare_mass_comparison_bar_plot} extends the cost comparison performed earlier for the KL method alone to all four method-kernel combinations, at the same reference level $|am_{\mathrm{bare}}^{\mathrm{critical}}| \approx 4\times10^{-4}$. For the Chebyshev method, only data beyond the initial oscillatory regime are considered ($N \geq 210$ for the Wilson kernel and $N \geq 45$ for
            the Brillouin). Within that regime, small values of $|am_{\mathrm{bare}}^{\mathrm{critical}}|$ may occur as spurious artifacts of a poor approximation to the matrix sign function rather than as a genuine reflection of chiral symmetry. Subject to these caveats, the KL-Brillouin combination reaches the reference level at roughly $20\%$ lower computational cost than the Chebyshev-Wilson combination, confirming the advantage of combining the KL rational approximation with the Brillouin kernel.
    
    
    \section{Conclusion}
    \label{sec:conclusions}
    
        In this paper we introduced and demonstrated the applicability of the diagonal Kenney-Laub (KL) rational approximation for the overlap operator in Lattice QCD.  We further demonstrated the usefulness of expressing the KL approximation via its partial fraction decomposition, which is straightforward to implement, with coefficients available in closed form and independent of spectral information. We showed that the approximation converges monotonically, improving in a systematic way with the diagonal KL order $n$.  The partial fraction form enabled the use of multi-shift conjugate gradient solvers, thus reducing computational costs and making higher-order approximations more accessible.
    
        The KL method paired with the Brillouin kernel was found to require fewer computational resources than the other method-kernel combinations considered, at least in terms of the resulting critical bare mass of the approximate overlap operator. We believe that these benefits warrant the effort for implementing the Brillouin operator, making the ``Brillouin plus KL-approximation'' combination both practical and effective in Lattice QCD computations.  Furthermore, the monotonic improvement of the KL method compared to the oscillatory convergence of the Chebyshev approximation significantly reduces the tuning effort required for choosing the approximation order $n$ that yields a given target precision.
    
    
    \section*{Acknowledgements}

        The authors gratefully acknowledge computing time on the supercomputer JURECA-DC at Forschungszentrum J\"ulich under grant no.~\texttt{overlapreloaded}~\cite{jureca-dc-2021}. Standard compute nodes feature two AMD EPYC 7742 processors (64 cores each at 2.25 GHz). Forward application studies of the overlap operator utilized single nodes, while physical calculations consistently employed 8 nodes for more reliable wall-clock time measurements.

        S.\,G.\ and G.\,K.\ acknowledge support by the PulseQCD, DeNuTra, HyperON, and Baryon8 projects (EXCELLENCE/0524/0269, EXCELLENCE/0524/0455, VISION ERC-PATH 2/0524/0001, and POSTDOC/0524/0001) co-financed by the European Regional Development Fund and the Republic of Cyprus through the Research and Innovation Foundation. G.\,K.\ acknowledges partial support from AQTIVATE, which received funding from the European Union's research and innovation program under the Marie Sklodowska-Curie Doctoral Networks action (Grant Agreement No.~101072344).
    
        
    
    \addcontentsline{toc}{section}{Appendices}
    
    \appendix


        \section{Implementation Details}
        \label{app:implementation}

            \subsection*{} 
            \vspace{-2em}  

            \subsubsection*{Massive overlap operator inversion}

                For both the KL rational and Chebyshev polynomial approximation methods, physical quantities were calculated by solving the massive-overlap equation $a D^{\mathrm{ov}}_m x = b$ using the Conjugate Gradient method applied to the Normal Equations for the Residual (CGNR):
                \begin{gather}
                    \Rightarrow \left( a D^{\mathrm{ov}}_m \right)^{\dagger} a D^{\mathrm{ov}}_m x = \left( a D^{\mathrm{ov}}_m \right)^{\dagger} b \equiv b' \; ,
                    \label{eq:CGNR_massive_overlap}
                \end{gather}
                where the Hermitian conjugate satisfies: $\left( a D^{\mathrm{ov}}_m \right)^{\dagger} = \gamma_5 a D^{\mathrm{ov}}_m \gamma_5$.

                The CGNR method was chosen for its guaranteed convergence properties. While this approach approximately doubles the computational cost due to the squared operator structure, it ensures convergence through the Hermiticity and positive-definiteness of $\left( a D^{\mathrm{ov}}_m \right)^{\dagger} a D^{\mathrm{ov}}_m$ for $m \neq 0$.

                A fixed stopping criterion based on the relative residual was imposed for both approximation methods:
                \begin{gather}
                    \frac{||\vec{r}||^2}{||\vec{b'}||^2} < \epsilon^2_{\text{CG}} \quad \text{with} \quad \epsilon_{\text{CG}} = 10^{-6} \; ,
                    \label{eq:CG_precision}
                \end{gather}
                where $\vec{r} = b' - \left( a D^{\mathrm{ov}}_m \right)^{\dagger} a D^{\mathrm{ov}}_m x$ is the residual vector of the modified problem. This choice provides sufficient precision for the physical observables studied while maintaining computational efficiency. Indeed, decreasing $\epsilon_{\text{CG}}$ to $10^{-7}$ provided no significant increase in precision for significantly greater computational cost.
    
                Lastly, for the inversion comparisons in Table~\ref{tab:partial_Vs_single_fraction_invert}, CGNR was applied to the single-fraction (SF) formulation as in Eq.~\ref{eq:CGNR_massive_overlap}, while for the ``multiply-up trick'' (MU) it was adjusted accordingly for Eq.~\ref{eq:multiply_up_trick}. The precision specified in Eq.~\ref{eq:CG_precision} was used for all results, with the exception of the Wilson kernel in the MU formulation, where higher precision was required (starting from $\epsilon_{\text{CG}} = 10^{-7}$ for $n=1$) to ensure consistent results across all formulations. This increased precision requirement can be attributed to the steeply increasing condition number of the polynomial operator on the left-hand side of Eq.~\ref{eq:multiply_up_trick} with increasing $n$, especially for the Wilson kernel, in contrast to the more stable rational SF and PF expressions.
                
            \subsubsection*{Multi-Shift solver and stopping criteria}
                            
                The matrix sign function approximation using the partial fraction expression of Eq.~\ref{eq:KL_sign_function_partial_fraction} requires solving multiple shifted systems simultaneously. An MSCG solver was employed to solve systems of the form:
                \begin{gather}
                    \left( \mathbb{X}^2 + \sigma_i \right) x = b
                \end{gather}
                where $\mathbb{X}$ is defined in Eq.~\ref{eq:gamma5_kernel} and the shifts $\sigma_i$ are given by Eq.~\ref{eq:KL_partia_fraction_shifts}. The Hermiticity of $\mathbb{X}^2$ and strict positivity of all shifts ($\sigma_i > 0$) guarantee convergence without additional transformations.
                
                A stopping criterion based again on the relative residual was used:
                \begin{gather}
                    \frac{||\vec{r}||^2}{||\vec{b}||^2} < \epsilon^2_{\text{MSCG}}, \quad \text{with} \quad \epsilon_{\text{MSCG}} \le 10^{-6} \; ,
                    \label{eq:MSCG_precision}
                \end{gather}
                depending on the resources available at the time. We found no significant difference in the PCAC mass obtained when using $\epsilon_{\text{MSCG}}=10^{-6}$ compared to $\epsilon_{\text{MSCG}}=10^{-7}$.  

                Lastly, for the SF results in Table~\ref{tab:partial_Vs_single_fraction}, the inversions of the polynomial $Q_{nn}(\mathbb{X}^2)$ (defined in Eq.~\ref{eq:KL_sign_function_single_fraction}) in the form $Q_{nn}(\mathbb{X}^2) x = b$ used the same stopping criterion and precision as in Eq.~\ref{eq:MSCG_precision}. Since $Q_{nn}(\mathbb{X}^2)$ is Hermitian and strictly positive-definite due to its constant term, a standard CG method was applied directly.

            \subsubsection*{Spectral bounds via Lanczos algorithm}
                
                As discussed in Appendix~\ref{app:Chebyshev}, the Chebyshev polynomial approximation of the matrix sign function requires estimating the extremal eigenvalues of $\mathbb{X}^2$. These were calculated using the Lanczos algorithm, which iteratively constructs a tridiagonal matrix whose eigenvalues (Ritz values) converge to the extrema of the $\mathbb{X}^2$ spectrum. The algorithm terminates when the relative change in the condition number $\kappa_{\mathbb{X}^2}^i = (\lambda^{\max}_{\mathbb{X}^2})^i/(\lambda^{\min}_{\mathbb{X}^2})^i$ between successive iterations satisfies:
                \begin{gather}
                    2\frac{|\kappa_{\mathbb{X}^2}^i - \kappa_{\mathbb{X}^2}^{i-1}|}{|\kappa_{\mathbb{X}^2}^i + \kappa_{\mathbb{X}^2}^{i-1}|} < \epsilon_{\text{Lanczos}}
                    \label{eq:Lanczos_precision}
                \end{gather}
                ensuring stable convergence of both extremal eigenvalues. The precision chosen was $\epsilon_{\text{Lanczos}} = 10^{-10}$; testing smaller precision values showed no significant improvement in accuracy for a substantially greater computational cost.
                
                The Lanczos algorithm typically overestimates the smallest eigenvalue and underestimates the largest one (in absolute magnitude). To account for these systematic deviations, we construct the Chebyshev approximation over a conservatively widened range: the lower bound is reduced by 50\% and the upper bound is increased by 10\% relative to the Lanczos estimates. This ensures that all eigenvalues of the rescaled operator (see Eq.~\ref{eq:matrix_linear_transformation}) remain within the $[-1,1]$ interval, preventing divergence in the Chebyshev approximation (cf.\,right panel of Fig.~\ref{fig:scalar_sign_function_approximations}).

        
        \section{Brillouin Improvement to the Wilson-Dirac Operator}
        \label{app:Brillouin}

            \begin{figure*}[ht]
                \centering
                \begin{minipage}{0.49\textwidth}
                    \centering
                    \includegraphics[scale=.95]{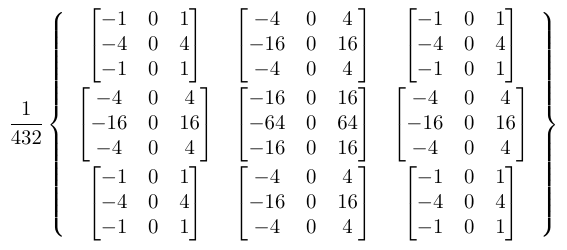}
                \end{minipage}%
                \begin{minipage}{0.49\textwidth}
                    \centering
                    \includegraphics[scale=.95]{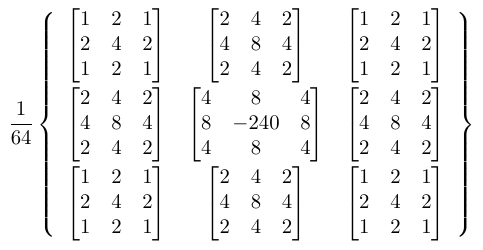}
                \end{minipage}

                \caption{Stencil representations of the two ingredients of the Brillouin operator in four dimensions. Left: the x-direction component of the covariant derivative, $\nabla_{\mu=1}$. Right: the Brillouin discretization of the gauged Laplacian, $\triangle$.}                
                \label{fig:Brillouin_operator_stencils}
            \end{figure*}
        
            The Brillouin operator is a parameter-free improvement of the Wilson-Dirac operator, designed to reduce rotational symmetry breaking near the edges of the Brillouin zone, hence its name.

            Let the general form of a discretized, doubler-free Dirac operator be:
            \begin{gather}
                D(x, y) = \sum_{\mu=1}^4 \gamma_\mu \nabla_\mu(x, y) - \frac{a}{2} \triangle(x, y) + m_0 \delta_{x, y} \nonumber\\
                -\frac{c_{\mathrm{SW}}}{2} \sum_{\mu<\nu} \sigma_{\mu \nu} F_{\mu \nu} \delta_{x, y} \; ,
                \label{eq:Dirac_operator_structure}
            \end{gather}
            where $\nabla_\mu$ and $\triangle$ represent the discretizations of the covariant derivative and gauged Laplacian, respectively. In this work, no clover-term improvement is applied ($c_\mathrm{SW} = 0$). As introduced in~\cite{Durr:2010ch}, the Brillouin operator in four dimensions employs a 64-point stencil for the ``isotropic'' covariant derivative and an 81-point stencil for the ``Brillouin'' gauged Laplacian, compared to the 2-point and 9-point stencils of the Wilson operator. Fig.~\ref{fig:Brillouin_operator_stencils} illustrates these stencils: the left panel shows the covariant derivative in the $x$-direction, while the right panel displays the complete Laplacian stencil.

            By incorporating contributions from an extended set of neighboring lattice points, the Brillouin operator achieves several improvements over the Wilson operator:
            
                \emph{Enhanced Isotropy:} The extended stencils improve the rotational symmetry of the discretization, reducing lattice artifacts associated with preferred directions (e.g., in the dispersion relations, see below).
    
                \emph{Improved Eigenvalue Spectrum:} The Brillouin operator exhibits a more ``shifted unitary'' eigenvalue spectrum compared to the Wilson kernel (see Fig.~5 in~\cite{Durr:2017wfi}), with spectral properties that more closely approach the ideal form $a \lambda = \rho \left( 1 - e^{-i \phi} \right)$, $\phi \in (-\pi, \pi]$, required by the Ginsparg-Wilson relation.
    
                \emph{Better Conditioning:} For a given gauge field configuration, the Brillouin kernel typically yields a smaller condition number for $\mathbb{X}^2$ compared to the Wilson case. 
                Here we define the condition number $\kappa_{\mathbb{X}^2} = \lambda^{\max}_{\mathbb{X}^2}/\lambda^{\min}_{\mathbb{X}^2}$, with $\mathbb{X}$ as defined in Eq.~\ref{eq:gamma5_kernel}.
                This improvement comes primarily from a reduced maximum eigenvalue, with $\lambda^{\max}_{\mathbb{X}^2} \simeq (2-\rho)^2$ for Brillouin~\cite{Durr:2010ch} compared to $\lambda^{\max}_{\mathbb{X}^2} \simeq (8-\rho)^2$ for Wilson~\cite{Neuberger:1999pz} in the free-field case.
    
                \emph{Improved Dispersion Relation:} The resulting dispersion relation more closely approximates that of the continuum theory~\cite{Durr:2012dw, Durr:2017wfi}.

            However, the Brillouin operator has several limitations. Due to its extended stencils, it requires a higher flop count per site and is more complex to implement than the Wilson operator~\cite{Ikeda:2009mv, Durr:2021iff}. Additionally, it provides only modest improvement to $\mathcal{O}(a)$ discretization errors, as its primary design focus is enhanced isotropy rather than Symanzik improvement.
        
            A notable feature that distinguishes the Brillouin operator from other chirally-improved Dirac operators in the literature is the absence of tunable parameters, eliminating uncertainties associated with parameter optimization and ensuring consistent behavior across different lattice setups.
            
            These properties make the Brillouin operator well-suited for use as a kernel in the overlap operator procedure, offering enhanced chiral properties without requiring parameter tuning.
            
        
        \section{Kenney-Laub Rational Approximation to the Matrix Sign Function}
        \label{app:KL}

            In~\cite{Durr:2017wfi}, we presented a detailed exposition of the Kenney-Laub family of iterations for approximating the matrix sign function. We focused particularly on the diagonal elements~\cite{kenney1994hyperbolic-364} of this family of iterations due to their advantageous properties, including monotonic growth and global stability. This appendix briefly summarizes the key concepts and formulas relevant to the calculations in this study. For a comprehensive treatment, see~\cite{KenneyLaub:1991, higham2008functions}.

            \begin{figure*}[ht]
                \centering
                \begin{minipage}{0.49\textwidth}
                    \centering
                    \includegraphics[width=\linewidth]{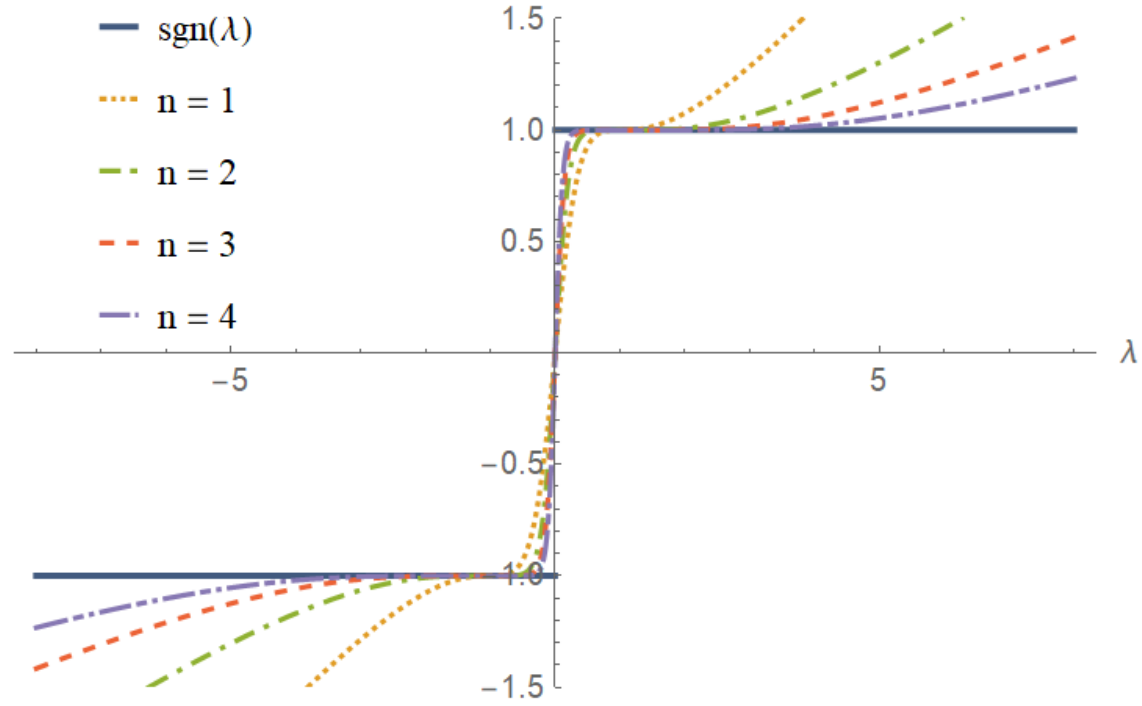}
                \end{minipage}
                \hfill
                \begin{minipage}{0.49\textwidth}
                    \centering
                    \includegraphics[width=\linewidth]{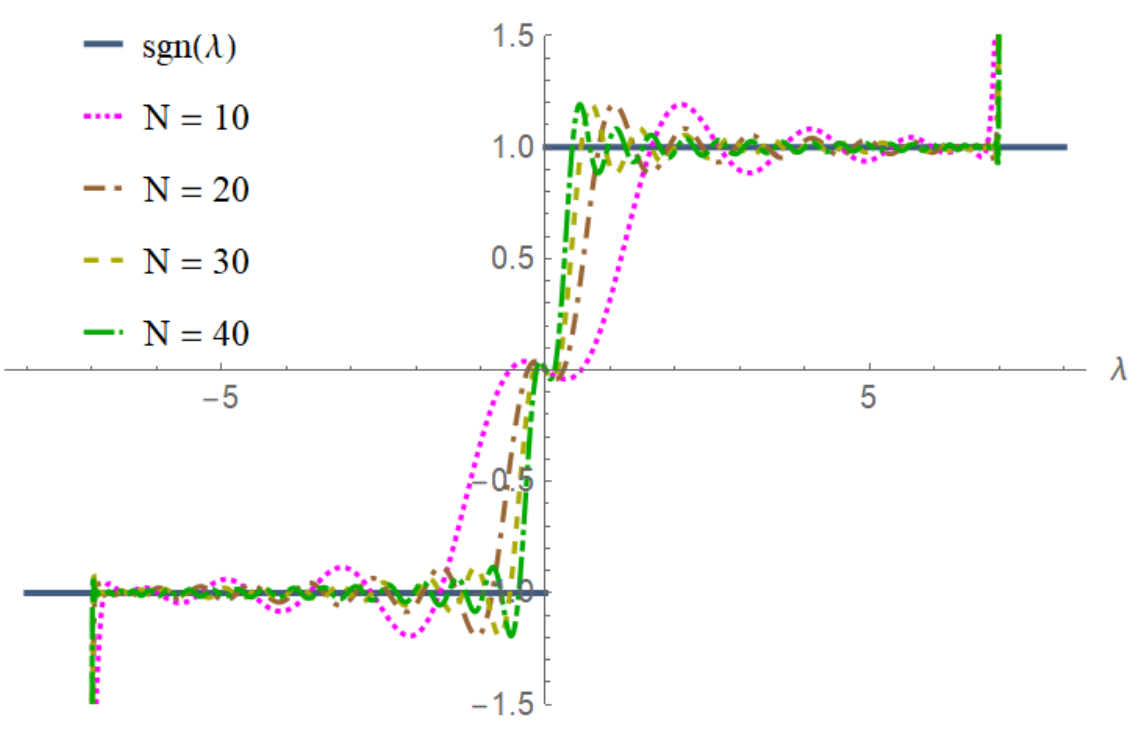}
                \end{minipage}

                \caption{Exact scalar sign function (solid blue line) compared with approximations (dashed lines) using the Kenney-Laub rational method for diagonal KL orders $n = 1, 2, 3, 4$ (left), and the Chebyshev polynomial method for polynomial orders $N = 10, 20, 30, 40$ (right). For the latter case, $\lambda_{\mathbb{H}^2}^{\min} = 10^{-6}$ and $\lambda_{\mathbb{H}^2}^{\max} = 49$ as defined for Eq.~\ref{eq:rescaled_scalar_func}, thus the range of approximation is $\lambda \in [-7.0, -0.001] \cup [0.001, 7.0]$.}
                \label{fig:scalar_sign_function_approximations}
            \end{figure*}

            Consider the matrix $\mathbb{H}$ that is supposed to be Hermitian with no zero eigenvalues\footnote{More generally, $\mathbb{H}$ need only be square with no purely imaginary eigenvalues, but Hermiticity is required in this work.}. The k-th recursion of the $(m,n)$ Kenney-Laub (KL) iterate $f_{mn}$ of $\mathbb{H}$ is defined as:
            \begin{align}
                \mathbb{H}_{k+1} = f_{mn}&\!\left( \mathbb{H}_k\right) \; ,
                \quad \mathbb{H}_0 = \mathbb{H} \\
                \text{with} \quad
                f_{mn}\left(\mathbb{H}_k\right) &= \mathbb{H}_k \frac{p_{mn}\left(\mathbb{I} - \mathbb{H}_k^2\right)}{q_{mn}\left(\mathbb{I} - \mathbb{H}_k^2\right)} \; ,
            \end{align}
            where $\mathbb{I}$ is the identity matrix, and the ratio -- in scalar terms -- $p_{mn}(t) / q_{mn}(t)$ with $t = 1 - x^2$ is the $(m,n)$ Pad\'{e} approximant to $h(t) = (1 - t)^{-1/2}$. Here, $p_{mn}$ is a polynomial of order $m$ in $t$ (or order $2m$ in $x$), while $q_{mn}$ is of order $n$ in $t$ (or $2n$ in $x$). The coefficients in both the numerator and denominator polynomials of the Pad\'{e} approximant depend on the degree of the other polynomial, hence the notation with both indices.
            
            It can be shown that an infinite sequence of recursions converges to the complex matrix sign function:
            \begin{align}
                \textrm{sgn}\!\left[ \mathbb{H} \right] = \lim_{k \to \infty} \mathbb{H}_k \; .
            \end{align}

            This work focuses specifically on the diagonal elements ($m=n$) for several reasons. They exhibit global convergence properties, ensuring reliable behavior regardless of the condition number of the argument matrix. Additionally, they demonstrate monotonic growth, providing stability and predictability in calculations. They satisfy a rather simple convergence error formula for the $k$-th recursion:
            \begin{align}
                \frac{\textrm{sgn}\!\left[ \mathbb{H} \right] - \mathbb{H}_k}{\textrm{sgn}\!\left[ \mathbb{H} \right] + \mathbb{H}_k} = \left( \frac{\textrm{sgn}\!\left[ \mathbb{H} \right] - \mathbb{H}_0}{\textrm{sgn}\!\left[ \mathbb{H} \right] + \mathbb{H}_0} \right)^{(2n+1)^k} \; .
                \label{eq:KL_convergence_error}
            \end{align}
            Furthermore, the diagonal elements satisfy useful group-like properties
            \footnote{This property follows from Eq.~14 in~\cite{Durr:2017wfi} with $l = 2n+1$.}:
            \begin{align*}
                f_{n'n'}\left( f_{nn}\left(\mathbb{H}_k\right) \right) = f_{n''n''}\left( \mathbb{H}_k \right) \; ,
            \end{align*}
            with $n''=(2n+1)\times(2n'+1)$, allowing multiple recursions of lower-order diagonal elements to be mapped to single recursions of higher-order elements with known analytical expressions.

            For practical reasons, we calculate only a single recursion of the diagonal elements for approximating the matrix sign function:
            \begin{align}
                \textrm{sgn}\!\left[ \mathbb{H} \right] \approx f_{nn}\left(\mathbb{H}\right) \; ,
            \end{align}
            and control the convergence with the overall approximation order $n$ (dubbed $n''$ above). Throughout this work, we refer to $n$ as the ``diagonal KL order''.

            General expressions for the Pad\'{e} approximant polynomials of the diagonal elements are\footnote{The expressions can be straightforwardly derived with a binomial expansion of Eq.~11 in~\cite{Durr:2017wfi} again with $l = 2n+1$.}:
            \begin{align}
                p_{nn}(\mathbb{I} - \mathbb{H}^2) &= \sum_{i=0}^{n} \binom{2n+1}{2i+1} ( \mathbb{H}^{2} )^i, \\
                q_{nn}(\mathbb{I} - \mathbb{H}^2) &= \sum_{i=0}^{n} \binom{2n+1}{2i} ( \mathbb{H}^{2} )^i,
            \end{align}
            which exhibit symmetric coefficients. Thus, a general expression for diagonal elements is:
            \begin{align}
                f_{nn}(\mathbb{H}) = \mathbb{H} \frac{ \sum_{i=0}^{n} \binom{2n+1}{2i+1} ( \mathbb{H}^{2} )^i }{ \sum_{i=0}^{n} \binom{2n+1}{2i} ( \mathbb{H}^{2} )^i } \; .
                \label{eq:KL_iterates_single_fraction}
            \end{align}

            The left panel of Fig.~\ref{fig:scalar_sign_function_approximations} shows the scalar KL rational approximation to the sign function using the diagonal KL iterates of Eq.~\ref{eq:KL_iterates_single_fraction} (in scalar form) for several diagonal KL orders compared to the exact sign function. The monotonicity of the diagonal KL iterates is apparent, along with their rapid convergence to the exact sign function with increasing diagonal KL order $n$.
            
            \subsection*{Partial fraction decomposition}

                Another particularly valuable property of the diagonal KL iterates is that they admit a partial fraction (PF) decomposition with a convenient form for numerical implementation%
                \footnote{
                    This form is equivalent to Eq.~17 of~\cite{Durr:2017wfi} after an index shift and a slight reorganization of the terms.
                }:
                \begin{align}
                    f_{nn}\left(\mathbb{H}\right) = \mathbb{H} \left( c_0 + \sum_{i=1}^{n} \frac{c_i}{ \mathbb{H}^2 + \sigma_i } \right) \; ,
                    \label{eq:KL_sign_function_partial_fraction}
                \end{align}
                with constant numerators $c_i$ and denominators involving only $\mathbb{H}^2$ and scalar shifts. The coefficients are:
                \begin{gather}
                    c_0 = \frac{1}{2 n + 1} , \quad 
                    c_i = \frac{2 c_0}{\cos^2\!\left[ \frac{(2i - 1) \pi}{4 n + 2} \right]} \; ,
                    \label{eq:KL_partia_fraction_const}
                \end{gather}
                and the shifts:
                \begin{gather}
                    \sigma_i = \tan^2\!\left[ \frac{(2i - 1) \pi}{4 n + 2} \right] \; ,
                    \label{eq:KL_partia_fraction_shifts}
                \end{gather}
                for $i = 1, \ldots, n$, with $0 < c_0 < c_1 < c_2 < \ldots < c_n$ and $0 < \sigma_1 < \sigma_2 < \ldots < \sigma_n$.

                This form is particularly well-suited for Multi-Shift Conjugate Gradient (MSCG) solvers: since all shifts $\sigma_i$ are strictly positive and $\mathbb{H}^2$ is Hermitian, the systems
                \begin{gather}
                    \left( \mathbb{H}^2 + \sigma_i \right) x = b
                \end{gather}
                can be solved simultaneously without additional transformations, with the total computational cost dominated by the smallest (slowest-converging) shift.
        

        \section{Chebyshev Polynomial Approximation to the Matrix Sign Function}
        \label{app:Chebyshev}
            
            This appendix summarizes the use of Chebyshev polynomials for approximating functions of matrices, focusing specifically on the inverse square root approximation necessary for constructing the matrix sign function of a Hermitian matrix. For a comprehensive treatment, see~\cite{higham2008functions, press2007numerical}.
        
            \subsection*{Chebyshev polynomial expansion of a matrix function}
        
                Let $\mathbb{H}$ be a Hermitian matrix with eigenvalues in the interval $[\lambda^{\mathrm{min}}, \lambda^{\mathrm{max}}] \subseteq \mathbb{R}$, where $\lambda^{\mathrm{min}} < \lambda^{\mathrm{max}}$, and let $f(x)$ be a well-defined real-valued function in this domain.
        
                The scalar function $f(x)$ can be expanded in Chebyshev polynomials of the first kind, $T_k(\tilde{x})$, where $\tilde{x} \in [-1,1]$ and $k = 0, 1, \ldots$ is the polynomial degree. This expansion exploits the orthogonality of Chebyshev polynomials but requires rescaling the argument $x$ to the interval $[-1,1]$. Consequently, we define a rescaled function $\tilde{f}(\tilde{x}) = f(x)$ and expand it in Chebyshev polynomials.
        
                The matrix function $f(\mathbb{H})$ is approximated using Chebyshev polynomials $T_k(\tilde{\mathbb{H}})$ of the rescaled matrix $\tilde{\mathbb{H}}$ whose eigenvalues lie within $[-1,1]$. This is most easily achieved via a linear transformation:
                \begin{align}
                    \tilde{\mathbb{H}} = \frac{2\mathbb{H} - (\lambda^{\mathrm{max}} + \lambda^{\mathrm{min}}) \mathbb{I}}{\lambda^{\mathrm{max}} - \lambda^{\mathrm{min}}},
                \end{align}
                that maps the range of eigenvalues from $[\lambda^{\mathrm{min}}, \lambda^{\mathrm{max}}]$ to $[-1,1]$. Consistency requires applying the same transformation to the scalar argument of $f(x)$:
                \begin{gather}
                    \tilde{x} = \frac{2x - (\lambda^{\mathrm{max}} + \lambda^{\mathrm{min}})}{\lambda^{\mathrm{max}} - \lambda^{\mathrm{min}}}, \\
                    \Rightarrow x = \frac{1}{2}(\lambda^{\mathrm{max}} - \lambda^{\mathrm{min}}) \tilde{x} + \frac{1}{2}(\lambda^{\mathrm{max}} + \lambda^{\mathrm{min}}),
                \end{gather}
                allowing the redefinition of $f(x)$:
                \begin{gather}
                    \tilde{f}(\tilde{x}) \equiv f\!\left( \frac{1}{2}(\lambda^{\mathrm{max}} - \lambda^{\mathrm{min}}) \tilde{x} + \frac{1}{2}(\lambda^{\mathrm{max}} + \lambda^{\mathrm{min}}) \right) \; .
                \end{gather}
        
                The matrix function is then approximated as:
                \begin{align}
                    f(\mathbb{H}) \approx \sum_{n=0}^{N-1} c_n T_n(\tilde{\mathbb{H}}),
                    \label{eq:Chebyshev_general_expansion}
                \end{align}
                where the Chebyshev coefficients are:
                \begin{align}
                    c_n = \frac{2-\delta_{n0}}{N} \sum_{k=0}^{N-1} \tilde{f}(x_k) T_n(x_k),
                    \label{eq:Chebyshev_coefficients}
                \end{align}
                with $\tilde{f}(x_k)$ and $T_n(x_k)$ evaluated at the Chebyshev nodes:
                \begin{align}
                    x_k = \cos\left(\frac{(2k + 1)\pi}{2N}\right), \quad k = 0, 1, \ldots, N-1.
                \end{align}
                The expansion of Eq.~\ref{eq:Chebyshev_general_expansion} has polynomial degree $N-1$ in $\mathbb{H}$.

                A three-term recurrence relation is satisfied by the Chebyshev polynomials of the first kind:
                \begin{gather}
                    T_0(\tilde{\mathbb{H}}) = \mathbb{I},
                    \quad T_1(\tilde{\mathbb{H}}) = \tilde{\mathbb{H}}, \\ 
                    T_{k+1}(\tilde{\mathbb{H}}) = 2\tilde{\mathbb{H}} \, T_k(\tilde{\mathbb{H}}) - T_{k-1}(\tilde{\mathbb{H}}) \; .
                    \label{eq:Chebyshev_recursion_relation}
                \end{gather}
                The Clenshaw summation algorithm makes use of this recurrence relation to provide efficient evaluation of Eq.~\ref{eq:Chebyshev_general_expansion} using backward recurrence with intermediate matrices $\mathbb{L}_k$, starting from $\mathbb{L}_{N+1} = \mathbb{L}_{N+2} = 0$:
                \begin{align}
                    \mathbb{L}_k &= 2\tilde{\mathbb{H}} \mathbb{L}_{k+1} - \mathbb{L}_{k+2} + c_k \mathbb{I} \; .
                \end{align}
                Thus a more compact expression for Eq.~\ref{eq:Chebyshev_general_expansion} is:
                \begin{align}
                    f(\mathbb{H}) &= \tilde{\mathbb{H}} \mathbb{L}_1 - \mathbb{L}_2 + c_0 \mathbb{I} \; .
                \end{align}
            
            \medskip
         
            \subsection*{\texorpdfstring{Approximation of $\left( \mathbb{H}^2 \right)^{-\frac{1}{2}}$}{Approximation of H squared inverse square root}}
                
                For the matrix sign function approximation, we require the inverse square root of $\mathbb{H}^2$, see Eq~\ref{eq:matrix_sign_function_definition}. Since $\mathbb{H}$ is Hermitian, $\mathbb{H}^2$ is positive semi-definite. We further require that it has no zero eigenvalues. Let $\mathbb{Y} \equiv \mathbb{H}^2$ with eigenvalues in $[\lambda^{\mathrm{min}}_{\mathbb{H}^2}, \lambda^{\mathrm{max}}_{\mathbb{H}^2}]$, where $0 < \lambda^{\mathrm{min}}_{\mathbb{H}^2} < \lambda^{\mathrm{max}}_{\mathbb{H}^2}$.
                
                The target function is $f(y)=y^{-1/2}$ for $y$ over the interval $[\lambda^{\mathrm{min}}_{\mathbb{H}^2}, \lambda^{\mathrm{max}}_{\mathbb{H}^2}]$. We map $\mathbb{Y}$ to $\tilde{\mathbb{Y}}$ with eigenvalues in $[-1,1]$:
                \begin{align}
                    \tilde{\mathbb{Y}} = \frac{2\mathbb{Y} - (\lambda^{\mathrm{max}}_{\mathbb{H}^2} + \lambda^{\mathrm{min}}_{\mathbb{H}^2}) \mathbb{I}}{\lambda^{\mathrm{max}}_{\mathbb{H}^2} - \lambda^{\mathrm{min}}_{\mathbb{H}^2}} \; , \label{eq:matrix_linear_transformation}
                \end{align}
                giving the rescaled function:
                \begin{align}
                    \tilde{f}(\tilde{y}) = \left[\frac{1}{2}(\lambda^{\mathrm{max}}_{\mathbb{H}^2} - \lambda^{\mathrm{min}}_{\mathbb{H}^2}) \tilde{y} + \frac{1}{2}(\lambda^{\mathrm{max}}_{\mathbb{H}^2} + \lambda^{\mathrm{min}}_{\mathbb{H}^2})\right]^{-1/2} \; .
                    \label{eq:rescaled_scalar_func}
                \end{align}
                
                The inverse square root is approximated as an expansion:
                \begin{align}
                    \left( \mathbb{H}^2 \right)^{-\frac{1}{2}} \approx \sum_{n=0}^{N-1} c_n T_n(\tilde{\mathbb{H}}^2) \; ,
                    \label{eq:inverse_square_root_expansion}
                \end{align}
                with coefficients $c_n$ computed using Eq.~\ref{eq:Chebyshev_coefficients}, or more conveniently using the Clenshaw algorithm:
                \begin{align}
                    \left( \mathbb{H}^2 \right)^{-\frac{1}{2}} = \tilde{\mathbb{H}}^2 \mathbb{L}_1 - \mathbb{L}_2 + c_0 \mathbb{I}  \; ,
                \end{align}
                from the backward recurrence ($\mathbb{L}_{N+1} = \mathbb{L}_{N+2} = 0$):
                \begin{align}
                    \mathbb{L}_k = 2\tilde{\mathbb{H}}^2 \mathbb{L}_{k+1} - \mathbb{L}_{k+2} + c_k \mathbb{I}.
                \end{align}
        
                The matrix sign function is then:
                \begin{align}
                    \text{sgn}[\mathbb{H}] = \mathbb{H} \left( \mathbb{H}^2 \right)^{-\frac{1}{2}} \approx \mathbb{H} \sum_{n=0}^{N-1} c_n T_n(\tilde{\mathbb{H}}^2) \; .
                    \label{eq:Chebyshev_sign_function_approximation}
                \end{align}
                The polynomial degree in $\mathbb{H}$ is $2(N-1)$ for Eq.~\ref{eq:inverse_square_root_expansion} and $2N-1$ for Eq.~\ref{eq:Chebyshev_sign_function_approximation}.
                
                Fig.~\ref{fig:scalar_sign_function_approximations} (right panel) shows the scalar sign function approximation for several values of $N$, compared to the exact scalar sign function, demonstrating the oscillatory character and rapid convergence of the Chebyshev method, particularly away from zero. The figure also illustrates the divergent behavior outside the approximation range of $y^{-1/2}$, $y \in [\lambda^{\mathrm{min}}_{\mathbb{H}^2}, \lambda^{\mathrm{max}}_{\mathbb{H}^2}]$, emphasizing the importance of safeguarding against possible imprecise eigenvalue estimation, see Appendix \ref{app:implementation}.


                \bibliography{references}

@article{Neuberger:1997fp, 
    year     = {1998}, 
    title    = {Exactly massless quarks on the lattice}, 
    author   = {Neuberger, Herbert}, 
    journal  = {Physics Letters B}, 
    issn     = {0370-2693}, 
    eprint   = {hep-lat/9707022}, 
    pages    = {141--144}, 
    number   = {1-2}, 
    volume   = {417}
}

@article{Neuberger:1998wv, 
    year     = {1998}, 
    title    = {More about exactly massless quarks on the lattice}, 
    author   = {Neuberger, Herbert}, 
    journal  = {Physics Letters B}, 
    issn     = {0370-2693}, 
    eprint   = {hep-lat/9801031}, 
    pages    = {353--355}, 
    number   = {3-4}, 
    volume   = {427}
}

@article{Neuberger:1998my, 
    year = {1998}, 
    title = {{A Practical Implementation of the Overlap Dirac Operator}}, 
    author = {Neuberger, Herbert}, 
    journal = {Physical Review Letters}, 
    issn = {0031-9007}, 
    eprint = {hep-lat/9806025}, 
    pages = {4060--4062}, 
    number = {19}, 
    volume = {81}, 
}

@article{Durr:2010ch,
    author = "Durr, Stephan and Koutsou, Giannis",
    title = "{Brillouin improvement for Wilson fermions}",
    eprint = "1012.3615",
    archivePrefix = "arXiv",
    primaryClass = "hep-lat",
    journal = "Phys. Rev. D",
    volume = "83",
    pages = "114512",
    year = "2011"
}

@article{Durr:2017wfi,
    author = "Durr, Stephan and Koutsou, Giannis",
    title = "{On the suitability of the Brillouin action as a kernel to the overlap procedure}",
    eprint = "1701.00726",
    journal = {},
    archivePrefix = "arXiv",
    primaryClass = "hep-lat",
    month = "1",
}

@article{Ginsparg:1981bj, 
    year     = {1982}, 
    title    = {A remnant of chiral symmetry on the lattice}, 
    author   = {Ginsparg, Paul H. and Wilson, Kenneth G.}, 
    journal  = {Physical Review D}, 
    issn     = {1550-7998}, 
    pages    = {2649--2657}, 
    number   = {10}, 
    volume   = {25}
}

@book{higham2008functions,
    title={Functions of Matrices: Theory and Computation},
    author={Higham, N.J.},
    isbn={9780898717778},
    year={2008},
    publisher={Society for Industrial and Applied Mathematics}
}

@article{KenneyLaub:1991, 
    year = {1991}, 
    title = {{Rational Iterative Methods for the Matrix Sign Function}}, 
    author = {Kenney, Charles and Laub, Alan J.}, 
    journal = {SIAM Journal on Matrix Analysis and Applications}, 
    issn = {0895-4798}, 
    pages = {273--291}, 
    number = {2}, 
    volume = {12},
}

@book{press2007numerical,
    title={Numerical Recipes 3rd Edition: The Art of Scientific Computing},
    author={Press, W.H.},
    isbn={9780521880688},
    lccn={2007062003},
    year={2007},
    publisher={Cambridge University Press}
}

@article{Edwards:1998yw, 
    year = {1999}, 
    title = {{A study of practical implementations of the overlap Dirac operator in four dimensions}}, 
    author = {Edwards, Robert G. and Heller, Urs M. and Narayanan, Rajamani}, 
    journal = {Nuclear Physics B}, 
    issn = {0550-3213}, 
    eprint = {hep-lat/9807017}, 
    pages = {457--471}, 
    number = {1-2}, 
    volume = {540}
}

@article{Jegerlehner:1996pm,
    author = "Jegerlehner, Beat",
    title = "{Krylov space solvers for shifted linear systems}",
    journal  = {},
    eprint = "hep-lat/9612014",
    reportNumber = "IUHET-353",
    month = "12",
}

@article{Frommer:1995ik, 
    year     = {1995}, 
    title    = {{Many} {Masses} {on} {one} {Stroke}: {Economic} {Computation} {of} {Quark} {Propagators}}, 
    author   = {{Frommer}, {Andreas} and {Nöckel}, {BERTOLD} and {Güsken}, {Stephan} and {Lippert}, {Thomas} and {Schilling}, {Klaus}}, 
    journal  = {International Journal of Modern Physics C}, 
    issn     = {0129-1831}, 
    eprint   = {hep-lat/9504020}, 
    pages    = {627--638}, 
    number   = {05}, 
    volume   = {6}
}

@article{Brannick:2014vda,
    author = {Brannick, James and Frommer, Andreas and Kahl, Karsten and Leder, Bj{\"o}rn and Rottmann, Matthias and Strebel, Artur},
    title = "{Multigrid Preconditioning for the Overlap Operator in Lattice QCD}",
    eprint = "1410.7170",
    archivePrefix = "arXiv",
    primaryClass = "hep-lat",
    journal = "Numer. Math.",
    volume = "132",
    number = "3",
    pages = "463--490",
    year = "2016"
}

@article{Ikeda:2009mv,
    author = "Ikeda, H. and Hashimoto, S.",
    editor = "Liu, Chuan and Zhu, Yu",
    title = "{O(a$^2$) improvement of the overlap-Dirac operator}",
    eprint = "0912.4119",
    archivePrefix = "arXiv",
    primaryClass = "hep-lat",
    journal = "PoS",
    volume = "LAT2009",
    pages = "082",
    year = "2009"
}

@article{vandenEshof:2002ms, 
    year     = {2002}, 
    title    = {{Numerical methods for the {QCD} overlap operator. I. Sign-function and error bounds}}, 
    author = {{J.~van den Eshof, A.~Frommer, T.~Lippert, K.~Schilling and H.~A.~van der Vorst}},
    journal  = {Computer Physics Communications}, 
    issn     = {0010-4655}, 
    eprint   = {hep-lat/0202025}, 
    pages    = {203--224}, 
    number   = {2}, 
    volume   = {146}, 
}

@inproceedings{Alexandru:2011sc,
    author = "Alexandru, Andrei and Lujan, Michael and Pelissier, Craig and Gamari, Ben and Lee, Frank X.",
    title = "{Efficient implementation of the overlap operator on multi-GPUs}",
    booktitle = "{2011 Symposium on Application Accelerators in High-Performance Computing (SAAHPC'11)}",
    eprint = "1106.4964",
    archivePrefix = "arXiv",
    primaryClass = "hep-lat",
    series = "IEEE Nucl.Sci.Symp.Conf.Rec.",
    pages = "123--130",
    year = "2011"
}

@article{Neuberger:1999pz, 
    year     = {1999}, 
    title    = {Bounds on the Wilson Dirac operator}, 
    author   = {Neuberger, Herbert}, 
    journal  = {Physical Review D}, 
    issn     = {1550-7998}, 
    eprint   = {hep-lat/9911004}, 
    pages    = {085015}, 
    number   = {8}, 
    volume   = {61}
}

@article{Durr:2021iff,
    author = "Durr, Stephan",
    title = "{Portable CPU implementation of Wilson, Brillouin and Susskind fermions in lattice QCD}",
    eprint = "2112.14640",
    archivePrefix = "arXiv",
    primaryClass = "hep-lat",
    journal = "Comput. Phys. Commun.",
    volume = "282",
    pages = "108555",
    year = "2023"
}

@article{Durr:2012dw,
    author = "Durr, Stephan and Koutsou, Giannis and Lippert, Thomas",
    title = "{Meson and Baryon dispersion relations with Brillouin fermions}",
    eprint = "1208.6270",
    archivePrefix = "arXiv",
    primaryClass = "hep-lat",
    journal = "Phys. Rev. D",
    volume = "86",
    pages = "114514",
    year = "2012"
}

@article{Boyle:2017jwu,
    author = {Boyle, Peter A. and Del Debbio, Luigi and J{\"u}ttner, Andreas and Khamseh, Ava and Sanfilippo, Francesco and Tsang, Justus Tobias},
    title = "{The decay constants ${\mathbf{f_D}}$ and ${\mathbf{f_{D_{s}}}}$ in the continuum limit of ${\mathbf{N_f=2+1}}$ domain wall lattice QCD}",
    eprint = "1701.02644",
    archivePrefix = "arXiv",
    primaryClass = "hep-lat",
    journal = "JHEP",
    number = "12",
    volume = "12",
    pages = "008",
    year = "2017"
}

@article{Luscher:1996sc,
    author = "Luscher, Martin and Sint, Stefan and Sommer, Rainer and Weisz, Peter",
    title = "{Chiral symmetry and O(a) improvement in lattice QCD}",
    eprint = "hep-lat/9605038",
    reportNumber = "DESY-96-086, CERN-TH-96-138, MPI-PHT-96-38",
    journal = "Nucl. Phys. B",
    volume = "478",
    pages = "365--400",
    year = "1996"
}

@article{Cundy:2010uq,
    author = "Cundy, Nigel and Kennedy, A. D. and Schafer, Andreas",
    title = "{A lattice Dirac operator for QCD with light dynamical quarks}",
    eprint = "1010.5629",
    archivePrefix = "arXiv",
    primaryClass = "hep-lat",
    journal = "Nucl. Phys. B",
    volume = "845",
    pages = "30--47",
    year = "2011"
}

@article{kenney1994hyperbolic-364, 
    year     = {1994}, 
    title    = {A hyperbolic tangent identity and the geometry of Padé sign function iterations}, 
    author   = {Kenney, C. S. and Laub, A. J.}, 
    journal  = {Numerical Algorithms}, 
    issn     = {1017-1398}, 
    pages    = {111--128}, 
    number   = {2}, 
    volume   = {7}
}

@inproceedings{Durr:2025cxn,
    author = {D{\"u}rr, Stephan and Gregoriou, Stylianos and Koutsou, Giannis},
    title = "{Diagonal Kenney-Laub Rational Approximation to the Overlap Dirac Operator}",
    booktitle = "{42th International Symposium on Lattice Field Theory}",
    eprint = "2512.20223",
    archivePrefix = "arXiv",
    primaryClass = "hep-lat",
    month = "12",
    year = "2025"
}

@article{APE:1987ehd,
    author = "Albanese, M. and others",
    collaboration = "APE",
    title = "{Glueball Masses and String Tension in Lattice QCD}",
    reportNumber = "ROM2F-87-014",
    journal = "Phys. Lett. B",
    volume = "192",
    pages = "163--169",
    year = "1987"
}

@article{DeGrand:2000tf,
    author = "DeGrand, Thomas A.",
    collaboration = "MILC",
    title = "{A Variant approach to the overlap action}",
    eprint = "hep-lat/0007046",
    reportNumber = "COLO-HEP-446",
    journal = "Phys. Rev. D",
    volume = "63",
    pages = "034503",
    year = "2000"
}

@article{jureca-dc-2021,
    author = {{J\"{u}lich Supercomputing Centre}},
    title = {{JURECA: Data Centric and Booster Modules implementing the Modular Supercomputing Architecture at J\"{u}lich Supercomputing Centre}},
    journal = {Journal of large-scale research facilities},
    number = {A182},
    volume = {7},
    url = {http://dx.% doi.org/10.17815/jlsrf-7-182},
    year = {2021}
}

@article{Luscher:1998pqa, 
    year     = {1998}, 
    title    = {Exact chiral symmetry on the lattice and the Ginsparg-Wilson relation}, 
    author   = {Lüscher, Martin}, 
    journal  = {Physics Letters B}, 
    issn     = {0370-2693}, 
    eprint   = {hep-lat/9802011}, 
    pages    = {342--345}, 
    number   = {3-4}, 
    volume   = {428} 
}

@article{DelDebbio:2003rn, 
    year     = {2003}, 
    title    = {Topological susceptibility from the overlap}, 
    author   = {Debbio, Luigi Del and Pica, Claudio}, 
    journal  = {Journal of High Energy Physics}, 
    eprint   = {hep-lat/0309145}, 
}

@article{Kennedy:2004tj, 
    year     = {2004}, 
    title    = {Approximation theory for matrices}, 
    author   = {Kennedy, A.D.}, 
    journal  = {Nuclear Physics B - Proceedings Supplements}, 
    issn     = {0920-5632}, 
    eprint   = {hep-lat/0402037}, 
    pages    = {107--116}, 
    volume   = {128}
}

@article{Bietenholz:2002ks,
    author = {Bietenholz, Wolfgang},
    title = {{Convergence rate and locality of improved overlap fermions}},
    eprint = {hep-lat/0204016},
    reportNumber = {HU-EP-02-12},
    journal = {Nucl. Phys. B},
    volume = {644},
    pages = {223--247},
    year = {2002},
}

@article{Cho:2015ffa,
    author = {Cho, Yong-Gwi and Hashimoto, Shoji and J{\"u}ttner, Andreas and Kaneko, Takashi and Marinkovic, Marina and Noaki, Jun-Ichi and Tsang, Justus Tobias},
    title = {{Improved lattice fermion action for heavy quarks}},
    eprint = {1504.01630},
    primaryClass = {hep-lat},
    reportNumber = {UTHEP-673, KEK-CP-321, CERN-PH-TH-2015-074},
    journal = {JHEP},
    pages = {72},
    year = {2015},
    archivePrefix = {arXiv},
    number   = {5}, 
    volume   = {2015}
}

\end{document}